\documentclass[10pt, letter, showpacs, titlepage, showkeys, floatfix, aps]{revtex4-1}
\usepackage{times}
\usepackage{hyperref}
\usepackage{amsmath}
\usepackage{amssymb}
\usepackage{graphicx}
\usepackage{xcolor} 
\usepackage{booktabs}
\usepackage{rotating}
\usepackage{bm}
\usepackage{float}
\usepackage{epstopdf}
\usepackage{comment}
\usepackage{tikz}
\usetikzlibrary{quantikz2}

\makeatletter

\newcommand{\Rmnum}[1]{\expandafter\@slowromancap\romannumeral #1@}
\makeatother

\usepackage{dirtytalk}
\usepackage{caption}
\usepackage{subcaption}

\begin{document}

\title{Exploring Entanglement and Parameter Sensitivity in QAOA through Quantum Fisher Information}

\author{Brian García Sarmina$^{1}$}
\email[E-mail:]{brian.garsar.6@gmail.com}

\author{Jorge Saavedra Benavides$^{1}$}
\email[E-mail:]{jsaavedrab2024@cic.ipn.mx}

\author{Guo-Hua Sun$^{1}$}
\email[E-mail:]{sunghdb@yahoo.com}

\author{Shi-Hai Dong$^{2,1}$}
\email[E-mail:]{dongsh2@yahoo.com}

\affiliation{$^1$ Centro de Investigaci\'{o}n en Computaci\'{o}n, Instituto Polit\'{e}cnico Nacional, UPALM, CDMX 07738, Mexico}
\affiliation{$^2$ Research Center for Quantum Physics, Huzhou University, Huzhou 313000, China}


\begin{abstract}
Quantum Fisher Information (QFI) can be used to quantify how sensitive a quantum state reacts to changes in its variational parameters, making it a natural diagnostic for algorithms such as the Quantum Approximate Optimization Algorithm (QAOA). We perform a systematic QFI analysis of QAOA for Max-Cut on cyclic and complete graphs with $N=4$-$10$ qubits. Two mixer families are studied, RX-only and hybrid RX-RY, with depths $p={2,4,6}$ and $p={3,6,9}$, respectively, and with up to three entanglement stages implemented through cyclic- or complete-entangling patterns. Complete graphs consistently yield larger QFI eigenvalues than cyclic graphs; none of the settings reaches the Heisenberg limit ($4N^{2}$), but several exceed the linear bound ($4N$). Introducing entanglement primarily redistributes QFI from diagonal to off-diagonal entries: non-entangled circuits maximize per-parameter (diagonal) sensitivity, whereas entangling layers increase the covariance fraction and thus cross-parameter correlations, with diminishing returns beyond the first stage. Leveraging these observations, we propose, as a proof of concept, a QFI‑Informed Mutation (QIm) heuristic that sets mutation probabilities and step sizes from the normalized diagonal QFI. On 7- and 10-qubit instances, QIm attains higher mean energies and lower variance than equal-probability and random-restart baselines over 100 runs, underscoring QFI as a lightweight, problem-aware preconditioner for QAOA and other variational quantum algorithms.
\end{abstract}

\maketitle

\section{Introduction}

Quantum computing has emerged as a transformative paradigm with the potential to solve certain optimization problems more efficiently than classical algorithms. Among near‑term candidates, the Quantum Approximate Optimization Algorithm (QAOA) \cite{farhi2014quantum} has attracted considerable interest for tackling combinatorial tasks that can be cast as unconstrained quadratic binary programs—e.g., Max-Cut, the traveling salesman problem, and portfolio optimization \cite{Harrigan2021quantum, Zhou2020QAOAapplications, Wang2018quantumapproximate}.

QAOA is a hybrid quantum-classical scheme \cite{choi2019tutorial}: a parameterized quantum circuit prepares a variational state, and a classical optimizer updates those parameters to minimize the expectation value of a problem Hamiltonian. The circuit alternates between (i) a problem Hamiltonian that encodes the objective function and (ii) a mixing Hamiltonian that drives transitions among computational basis states. Because the optimization landscape can be highly nonconvex, featuring local minima and barren plateaus \cite{mcclean2018barren, Wang2021noiseinduced}, standard gradient-based or gradient-free optimizers frequently struggle \cite{streif2019comparison, zhou2023qaoa, brandhofer2022benchmarking, Zhou2020Quantum}. Noise, decoherence, and hardware imperfections further complicate parameter tuning in the NISQ era \cite{Preskill2018NISQ, Guerreschi2019qaoa}.

To mitigate these challenges, several heuristics and learning-based strategies have been explored, from precomputed parameter schedules \cite{zhou2023qaoa} to reinforcement learning \cite{yao2021policygradient}. A complementary direction leverages tools from quantum information theory. In particular, the Quantum Fisher Information (QFI) \cite{Braunstein1994statistical, Helstrom1976quantum}—a metrological quantity that bounds achievable precision \cite{Paris2009, Toth2014quantum, Demkowicz2015quantummetrology}—has recently been adopted to diagnose and precondition variational circuits \cite{Beckey2020Variational, Tan2021Variational, zhang2024predicting, Stokes2020quantumnatural}. QFI measures how sensitively a quantum state changes under infinitesimal parameter variations, thereby revealing which parameters are informative and how strongly they are correlated.

Beyond sensitivity, QFI is useful for assessing robustness to noise and the role of entanglement in variational circuits \cite{Gentini2019Noise-resilient, huang2024quantum, koczor2022quantum, Meyer2021fisherinformation}. Properly engineered entangling patterns can alter QFI scaling and may alleviate barren plateaus \cite{cerezo2021cost, larocca2022theory}. However, entanglement can also redistribute information from diagonal (single-parameter) sensitivity to off-diagonal (cross-parameter) correlations, which affects how easily a classical optimizer can navigate the landscape.

In this work we systematically study QFI in QAOA for Max-Cut on cyclic and complete graphs with $N=4$–$10$ qubits. We compare two mixer families: an RX-only mixer and a hybrid RX-RY mixer. For each we vary depth ($p=\{2,4,6\}$ for RX, $p=\{3,6,9\}$ for RX-RY) and implement up to three entanglement stages using either cyclic or complete entangling patterns. Besides tracking eigenvalue bounds, we quantify cross-parameter coupling via the \emph{covariance fraction} $r=\sum_{i\neq j}|F_{ij}|/\sum_i F_{ii}$. Our results show that (i) complete graphs yield larger QFI eigenvalues and higher $r$ than cyclic graphs, (ii) no configuration reaches the Heisenberg limit $4N^{2}$, although several exceed the linear scaling $4N$, and (iii) the first entanglement stage contributes the largest QFI change; additional stages give diminishing returns and often reduce $r$ (RX) or make it non-monotonic (RX–RY/complete).

Finally, as a proof of concept, we introduce a simple QFI‑Informed Mutation heuristic (QIm) that sets mutation probabilities and step sizes from the normalized diagonal QFI. On 7- and 10-qubit Max-Cut instances, QIm outperforms equal-probability (nonQIm) and random‑restart (RR) baselines over 100 trials, achieving higher mean energies and lower variance. These findings highlight QFI as a lightweight, problem-aware preconditioner for QAOA and other variational quantum algorithms.

The remainder of the paper is organized as follows. Section II reviews QAOA and QFI. Section III introduces the Max-Cut instances and graph topologies. Section IV presents our QFI results and analysis. Section V concludes and outlines future directions.

\section{Preliminary insights into QAOA and Quantum Fisher Information}

In this section we outline the QAOA workflow for Max-Cut on cyclic and complete graphs, and introduce the Quantum Fisher Information (QFI) matrix, the main tool we used to quantify parameter sensitivities and correlations across different QAOA models.

\subsection{QAOA}

QAOA is a variational algorithm that alternates two Hamiltonians inside a parameterized circuit. The \emph{problem} (cost) Hamiltonian $H_{P}$ encodes the objective. For Max-Cut we use
\begin{equation}
  H_{P}=\sum_{(i,j)\in E} J_{ij} Z_i Z_j,\qquad J_{ij}=1 ,
\end{equation}
each edge $(i,j)$ contributes according to the parity of the qubits. Equivalently, the usual Max-Cut cost is $(1-Z_i Z_j)/2$ up to an overall shift. The interactions are implemented with $\mathrm{RZZ}$ gates.

The \emph{mixing} Hamiltonian $H_{M}$ drives transitions between computational basis states. We study two families:
\begin{itemize}
  \item \textbf{RX-only mixer}
  \begin{equation}
    H_{M}=\sum_i X_i .
  \end{equation}
  Its entangled variant inserts a CNOT-based stage acting on an entanglement set $E_s$ (defined by the chosen pattern-cyclic or complete):
  \begin{equation}
    H_{M}^{(\text{ent})}=\sum_{(n_i,n_j)\in E_s}\frac{1}{2}\bigl(I-Z_{n_i}\bigr)X_{n_j}+\sum_i X_i .
  \end{equation}
  \item \textbf{RX–RY mixer}
  \begin{equation}
    H_{M}=\sum_i X_i+\sum_i Y_i ,
  \end{equation}
  with the entangled version
  \begin{equation}
    H_{M}^{(\text{ent})}= \sum_i X_i + \sum_{(n_i,n_j)\in E_s}\frac{1}{2}\bigl(I-Z_{n_i}\bigr)X_{n_j} + \sum_i Y_i .
  \end{equation}
\end{itemize}

Then we used the $H_{P}$ and $H_{M}$ to form the phase $U_{P}(\gamma_{k})$ and mixing $U_{M}(\beta_{k})$ operators respectively, where the $p$-layer QAOA state is
\begin{equation}
  \ket{\psi(\boldsymbol{\gamma},\boldsymbol{\beta})}
  = \prod_{k=1}^{p} U_M(\beta_k)\, U_P(\gamma_k)\, \ket{+}^{\otimes N},
\end{equation}
with $U_P(\gamma)=e^{-i\gamma H_P}$, $U_M(\beta)=e^{-i\beta H_M}$, and $\ket{+}^{\otimes N}$ is the equal superposition initial state prepared via Hadamards on all $N$ qubits.

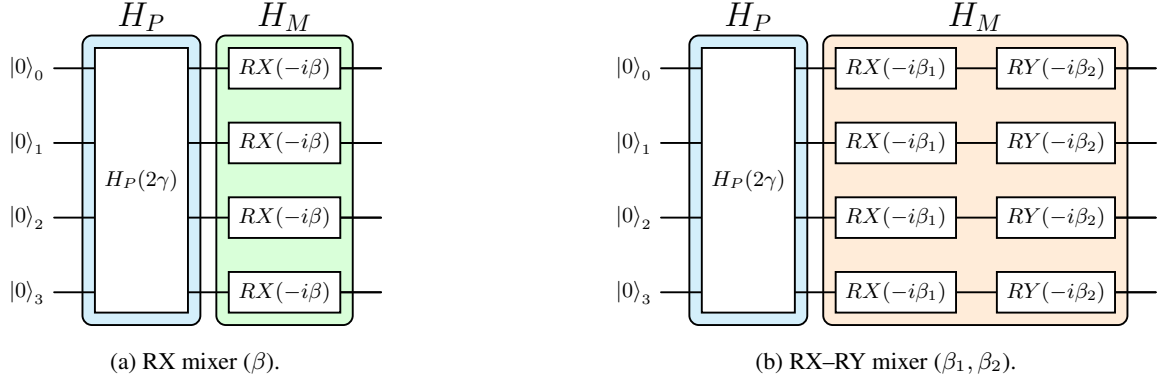
\begin{figure}
  \centering
  \begin{subfigure}[b]{0.49\textwidth}
    \centering
    \begin{tikzpicture}
      \node[scale=0.9]{
        \begin{quantikz}[row sep=0.5cm, column sep=0.6cm]
          \lstick{$\ket{0}_{0}$} &
            \gate[wires=4]{H_{P}(2\gamma)}
            \gategroup[wires=4,steps=1,
              style={rounded corners,fill=cyan!15,inner xsep=2pt,inner ysep=2pt},
              background]{\Large $H_{P}$} &
            \gate{RX(-i\beta)}
            \gategroup[wires=4,steps=1,
              style={rounded corners,fill=green!15,inner xsep=2pt,inner ysep=2pt},
              background]{\Large $H_{M}$} & \qw \\
          \lstick{$\ket{0}_{1}$} & & \gate{RX(-i\beta)} & \qw \\
          \lstick{$\ket{0}_{2}$} & & \gate{RX(-i\beta)} & \qw \\
          \lstick{$\ket{0}_{3}$} & & \gate{RX(-i\beta)} & \qw
        \end{quantikz}
      };
    \end{tikzpicture}
    \caption{RX mixer ($\beta$).}
    \label{fig:qaoa_rx}
  \end{subfigure}
  \hfill
  \begin{subfigure}[b]{0.49\textwidth}
    \centering
    \begin{tikzpicture}
      \node[scale=0.9]{
        \begin{quantikz}[row sep=0.5cm, column sep=0.6cm]
          \lstick{$\ket{0}_{0}$} &
            \gate[wires=4]{H_{P}(2\gamma)}
            \gategroup[wires=4,steps=1,
              style={rounded corners,fill=cyan!15,inner xsep=2pt,inner ysep=2pt},
              background]{\Large $H_{P}$} &
            \gate{RX(-i\beta_{1})}
              \gategroup[wires=4,steps=2,
                style={rounded corners,fill=orange!15,inner xsep=2pt,inner ysep=2pt},
                background]{\Large $H_{M}$} &
            \gate{RY(-i\beta_{2})} & \qw \\
          \lstick{$\ket{0}_{1}$} & & \gate{RX(-i\beta_{1})} & \gate{RY(-i\beta_{2})} & \qw \\
          \lstick{$\ket{0}_{2}$} & & \gate{RX(-i\beta_{1})} & \gate{RY(-i\beta_{2})} & \qw \\
          \lstick{$\ket{0}_{3}$} & & \gate{RX(-i\beta_{1})} & \gate{RY(-i\beta_{2})} & \qw
        \end{quantikz}
      };
    \end{tikzpicture}
    \caption{RX–RY mixer ($\beta_{1},\beta_{2}$).}
    \label{fig:qaoa_rxry}
  \end{subfigure}
  \caption{One-layer QAOA circuits on 4 qubits with RX-only and RX-RY mixers.}
  \label{fig:qaoa_rx_rxry_mixers}
\end{figure}

\begin{figure}
  \centering
  \begin{subfigure}[b]{0.43\textwidth}
    \centering
    \begin{tikzpicture}
      \node[scale=0.73]{
        \begin{quantikz}[row sep=0.5cm, column sep=0.6cm]
          \lstick{$\ket{0}_{0}$} &
            \gate[wires=4]{H_{P}(2\gamma)}
            \gategroup[wires=4,steps=1,
              style={rounded corners,fill=cyan!15,inner xsep=2pt,inner ysep=2pt},
              background]{\Large $H_{P}$} &
            \ctrl{1}
            \gategroup[wires=4,steps=4,
              style={rounded corners,fill=magenta!15,inner xsep=2pt,inner ysep=2pt},
              background]{\Large Entanglement} &
            \qw & \qw & \targ{} &
            \gate{RX(-i\beta)}
            \gategroup[wires=4,steps=1,
              style={rounded corners,fill=green!15,inner xsep=2pt,inner ysep=2pt},
              background]{\Large $H_{M}$} & \qw \\
          \lstick{$\ket{0}_{1}$} & & 
            \targ{} & \ctrl{1} & \qw & \qw &
            \gate{RX(-i\beta)} & \qw \\
          \lstick{$\ket{0}_{2}$} & & 
            \qw & \targ{} & \ctrl{1} & \qw &
            \gate{RX(-i\beta)} & \qw \\
          \lstick{$\ket{0}_{3}$} & & 
            \qw & \qw & \targ{} & \ctrl{-3} &
            \gate{RX(-i\beta)} & \qw
        \end{quantikz}
      };
    \end{tikzpicture}
    \caption{Cyclic entanglement + RX mixer ($\beta$).}
    \label{fig:qaoa_cyclic_rx}
  \end{subfigure}
  \hfill
  \begin{subfigure}[b]{0.56\textwidth}
    \centering
    \begin{tikzpicture}
      \node[scale=0.73]{
        \begin{quantikz}[row sep=0.5cm, column sep=0.6cm]
          \lstick{$\ket{0}_{0}$} &
            \gate[wires=4]{H_{P}(2\gamma)}
            \gategroup[wires=4,steps=1,
              style={rounded corners,fill=cyan!15,inner xsep=2pt,inner ysep=2pt},
              background]{\Large $H_{P}$} &
            \gate{RX(-i\beta_{1})}
            \gategroup[wires=4,steps=1,
              style={rounded corners,fill=orange!15,inner xsep=2pt,inner ysep=2pt},
              background]{\Large $H_{M}$} &
            \ctrl{1}
            \gategroup[wires=4,steps=6,
              style={rounded corners,fill=magenta!15,inner xsep=2pt,inner ysep=2pt},
              background]{\Large Entanglement} &
            \ctrl{2} & \ctrl{3} & \qw & \qw & \qw &
            \gate{RY(-i\beta_{2})}
            \gategroup[wires=4,steps=1,
              style={rounded corners,fill=orange!15,inner xsep=2pt,inner ysep=2pt},
              background]{\Large $H_{M}$} & \qw \\
          \lstick{$\ket{0}_{1}$} & &
            \gate{RX(-i\beta_{1})} & 
            \targ{} & \qw & \qw &
            \ctrl{1} & \ctrl{2} & \qw &
            \gate{RY(-i\beta_{2})} & \qw \\
          \lstick{$\ket{0}_{2}$} & &
            \gate{RX(-i\beta_{1})} &
            \qw & \targ{} & \qw &
            \targ{} & \qw & \ctrl{1} &
            \gate{RY(-i\beta_{2})} & \qw \\
          \lstick{$\ket{0}_{3}$} & &
            \gate{RX(-i\beta_{1})} &
            \qw & \qw & \targ{} &
            \qw & \targ{} & \targ{} &
            \gate{RY(-i\beta_{2})} & \qw
        \end{quantikz}
      };
    \end{tikzpicture}
    \caption{All-to-all (complete) entanglement between RX and RY ($\beta_{1},\beta_{2}$).}
    \label{fig:qaoa_complete_rxry}
  \end{subfigure}
  \caption{One-layer QAOA circuits on 4 qubits with different entanglement patterns.}
  \label{fig:qaoa_entanglement_variants}
\end{figure}
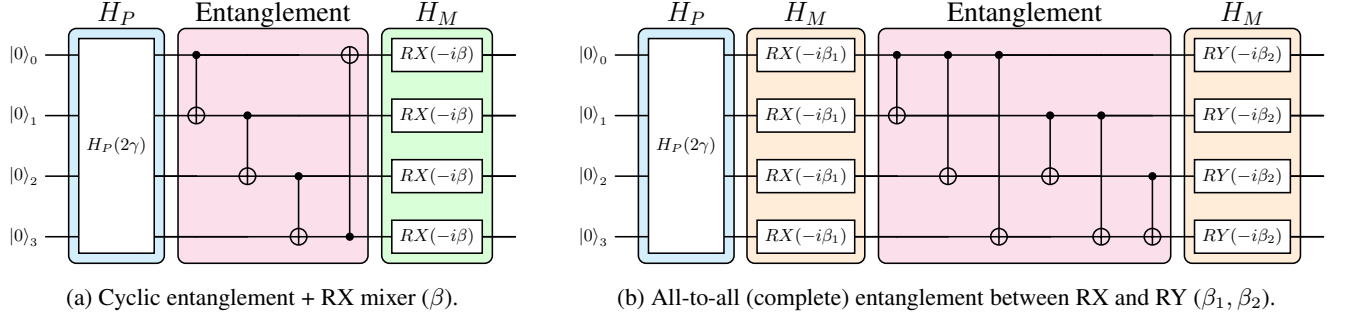

Figure \ref{fig:qaoa_rx_rxry_mixers} shows single-layer circuits for RX and RX-RY mixers, while Figure \ref{fig:qaoa_entanglement_variants} illustrates the two entanglement patterns (cyclic and complete) used to build the sets $E_{s}$.

\subsection{Quantum Fisher Information}

The Quantum Fisher Information (QFI) quantifies the sensitivity of a quantum state to infinitesimal changes in a set of control parameters, and thus provides a principled metric for assessing parameter informativeness in variational quantum algorithms. For a pure state $|\psi(\vec\theta)\rangle$, the QFI matrix is
\begin{equation}
    F_{ij}
    = 4\,\mathrm{Re} \left[
        \langle \partial_i \psi|\partial_j \psi \rangle
        - \langle \partial_i \psi|\psi \rangle \langle \psi|\partial_j \psi \rangle \right],
    \label{eq:qfim}
\end{equation}
with $|\partial_i \psi \rangle \equiv \partial_{\theta_i} |\psi(\vec\theta)\rangle$. The partial derivative $|\partial_i \psi \rangle$ was calculated using Linear Combination of Unitaries (LCU), details are given in Appendix A. In our work we compute the QFI from the quantum geometric tensor (QGT),
\begin{equation}
    \tau_{ij} = \tfrac{1}{4}F_{ij} + \tfrac{i}{2}\,\Omega_{ij},
\end{equation}
where $\Omega_{ij}$ is the Berry curvature.

Throughout, the parameter vector $\vec\theta$ comprises the QAOA angles: $\{\gamma_k,\beta_k\}$ for RX-only mixers and $\{\gamma_k,\beta^{(X)}_k,\beta^{(Y)}_k\}$ for RX–RY mixers. We assume a single parameter per gate stage (all RZZ gates within a phase layer share one $\gamma_k$, all RX gates in a mixer layer share one $\beta^{(X)}_k$, etc.), in contrast to multi-angle QAOA variants.

A convenient global summary of the QFIM (QFI matrix) is the \emph{sensitivity capacity},
\begin{equation}
\mathcal{F} \equiv \mathrm{Tr}\left[ F_{ij} \right] = \sum_{i=1}^m F_{ii},
\label{eq:sensitivity_capacity}
\end{equation}
with $m$ the number of variational parameters. In the multiparameter case, the quantum Cramér-Rao bound (QCRB) \cite{liuragy2016compatibility,liu2020quantumFisher} states that the covariance matrix $\mathrm{Cov}(\hat{\boldsymbol{\theta}})$ of any unbiased estimator $\hat{\boldsymbol{\theta}}$ satisfies:
\begin{equation}
\mathrm{Cov}(\hat{\boldsymbol{\theta}}) \geq (F_{ij})^{-1},
\label{eq:QCRB_multi}
\end{equation}
with $(F_{ij})^{-1}$ being the inverse of the QFIM. When all parameters are treated as unknown, the fundamental lower bound on the variance of $\theta_i$ is
\begin{equation}
\Delta^2 \theta_i \ge \bigl(F^{-1}_{ij}\bigr)_{ii}.
\label{eq:var_bound_multi}
\end{equation}

In the multiparameter setting, the off-diagonal entries of the QFIM play a crucial role because they quantify correlations between estimators of different variational angles. When the off-diagonal elements are $\neq 0$, the parameters are not statistically independent: information about $\theta_i$ is partially shared with, or contaminated by, information about other parameters. This coupling appears explicitly upon inverting the QFIM to obtain the multiparameter Cramér-Rao bound. In general one has $(F_{ij}^{-1})_{ii} \ge \frac{1}{F_{ii}}$ \cite{liu2020quantumFisher, liuragy2016compatibility}, so the attainable variance for $\theta_i$ in the joint-estimation problem is always worse (or at best equal) than the variance predicted by treating $\theta_i$ alone. Intuitively, the inversion redistributes the diagonal information through the off-diagonal blocks, reducing the \emph{effective} independent information available per parameter.

This degradation can be seen transparently in the two-parameter case, where
\begin{equation}
(F_{ij}^{-1})_{11} = \frac{F_{22}}{F_{11}F_{22}-F_{12}^2} = \frac{1}{F_{11}-\frac{F_{12}^2}{F_{22}}} \ .
\end{equation}
Hence, a nonzero correlation term $F_{12}$ subtracts from the effective Fisher information of $\theta_{1}$, enlarging its bound. As $|F_{12}|$ approaches $\sqrt{F_{11}F_{22}}$, the matrix becomes ill-conditioned and $(F_{ij}^{-1})_{ii}$ can grow rapidly, reflecting strong trade-offs between simultaneously estimating multiple angles \cite{liu2020quantumFisher, sidhu2020geometric}.

In our QAOA experiments, this mechanism elucidates why deeper or more entangling ansätze, which tend to populate the off-diagonal QFI elements \cite{Stokes2020quantumnatural, Meyer2021fisherinformation}, can result in an increased $\mathcal{F^{\text{-1}}} = \mathrm{Tr}[F^{-1}_{ij}]$ (\emph{total estimability}) even as $\mathcal{F}$ (sensitivity capacity) grows. The gain in raw sensitivity is effectively canceled out by stronger inter-parameter correlations, which renders the QFIM ill-conditioned and degrades global estimability.

By contrast, the simpler single-parameter bound
\begin{equation}
\Delta^2 \theta_i \ge \frac{1}{F_{ii}}
\label{eq:var_bound_single}
\end{equation}
applies only when $\theta_i$ is estimated while all other parameters are known (or effectively decoupled). Throughout this work we therefore interpret the diagonal entries $F_{ii}$ and the trace $\mathcal{F}$ as measures of local and global sensitivity capacity of the variational state to parameter variations.

A standard scalar measure of the total imprecision in simultaneous estimation is $\mathrm{Tr}[\mathrm{Cov}(\hat{\boldsymbol{\theta}})]$, which is bounded from below by $\mathcal{F^{\text{-1}}}$. For a noiseless single-parameter unitary $U_\theta = e^{-i \theta \Lambda}$ with generator $\Lambda$, the quantum Fisher information reduces to
\begin{equation}
F_Q = 4\,\mathrm{Var}(\Lambda) \le (\lambda_{\max} - \lambda_{\min})^2,
\end{equation}
where $\lambda_{\max}$ and $\lambda_{\min}$ are the largest and smallest eigenvalues of $\Lambda$.

Equivalently, for pure states
\begin{equation}
    F_{aa} = 4\bigl(\langle H_a^2\rangle - \langle H_a\rangle^2\bigr) = 4 \ \Delta H_{a}^{2}= 4 \ \mathrm{Var}(H_a),
\end{equation}
so $F_{aa}$ cannot exceed the squared spectral range of the corresponding generator $H_a$ \cite{gorecki2023heisenberg, mirkhalaf2016entanglement, wu2024molecular}. For example, $H_M=\sum_{i=1}^N X^{(i)}$ has spectrum in $[-N,N]$, where $N$ is the number of qubits, yielding $F_{aa}\le 4N^2$. In practice, however, the accessible QAOA states often behave closer to product states, giving shot-noise-like $\mathcal{O}(N)$ scaling for mixer parameters.

For RX-only QAOA, two parameters dominate: the cost angle $\gamma$ with generator $H_P$ and the mixer angle $\beta$ with generator $H_M=\sum_i X_i$.
\begin{itemize}
  \item \textbf{Cost ($\gamma$):} $F_{\gamma\gamma}=4\,\mathrm{Var}(H_P)$. For fully connected Ising costs $H_P=\sum_{i\neq j} Z_i Z_j$, the spectral gap scales as $O(N^2)$, so $F_{\gamma\gamma}$ could in principle reach $(O(N^2))^{2} = O(N^4)$ (super-Heisenberg \cite{rams2018limits, hou2021super, napolitano2011interaction}); empirically we do not observe this and adopt the conservative bound $F_{\gamma\gamma}\lesssim 4N^2$.
  \item \textbf{Mixer ($\beta$):} $F_{\beta\beta}=4\,\mathrm{Var}(H_M)\le 4N^2$, but non-commuting dynamics constrain the accessible states, typically yielding $F_{\beta\beta}\lesssim O(N)$.
\end{itemize}
A resulting coarse bound for RX-only models ($1L$ depth) is
\begin{equation}
    \mathcal{F} \lesssim 4\bigl(N^2 + N\bigr).
\end{equation}

For RX-RY mixers ($1L$ depth), an additional single-qubit generator $H_Y=\sum_i Y_i$ contributes analogously. Since $H_X$ and $H_Y$ do not commute, they cannot be simultaneously maximized; each one generally contributes at most $O(N)$ in the accessible regime. Thus,
\begin{equation}
    \mathcal{F} \lesssim 4\bigl(N^2 + 2N\bigr),
\end{equation}
with the quadratic term dominated by the cost layer and the mixers adding subleading linear contributions.

Beyond diagonal entries, off-diagonal elements capture parameter correlations. For pure states,
\begin{equation}
    F_{ij} = 4 \ \mathrm{Cov_{\psi}}(H_i,H_j), \quad i\neq j,
\end{equation}
with the $\mathrm{Cov}$ defined as $\text{Cov}_{\psi} (H_{i}, H_{j}) = \frac{1}{2} \langle H_{i} H_{j} + H_{j} H_{i}\rangle - \langle H_{i}\rangle \langle H_{j}\rangle$ for generalization \cite{liu2020quantumFisher}, and by Cauchy–Schwarz,
\begin{equation}
    |F_{ij}| \le 4\sqrt{\mathrm{Var}(H_i)\,\mathrm{Var}(H_j)}.
\end{equation}
Although off-diagonal terms do not change the asymptotic $N^2$ bound, they quantify cross-talk among parameters. To calculate this correlations between parameters (and architectures) we report the \emph{covariance fraction}
\begin{equation}
    r \equiv \frac{\sum_{i\neq j} |F_{ij}|}{\sum_i F_{ii}},
\end{equation}
a dimensionless measure ranging from $r \approx 0$ (diagonal dominance; nearly independent parameters) to $r \to m - 1$ for $m$ perfectly correlated parameters, indicating strong cross-parameter coupling. We use $r$ to compare entanglement patterns, depths, and mixer choices in the Results section.

\section{Problems}

We study Max-Cut instances on two canonical graph topologies, cyclic and complete, commonly used as benchmarks in the QAOA literature \cite{zhou2023qaoa,koczor2022quantum,guerreschi2019,commander2009maximum,kochenberger2013solving,laguna2009hybridizing}. These graphs provide clean test beds for assessing how entanglement structure, mixer choice, and circuit depth influence the Quantum Fisher Information (QFI).

To isolate QFI behavior from optimizer-induced effects, no classical parameter optimization is performed. Instead, for each configuration we draw 100 parameter vectors uniformly at random from $[0,2\pi)$ and compute the corresponding QFI matrices. Averaging these matrices yields a stable estimate of the global sensitivity landscape, while avoiding biases due to any particular optimizer trajectory. Consequently, we do not report Max-Cut solution qualities; analyzing the interplay between QFI profiles, optimization strategies, and performance is deferred to future work.

\begin{figure}[!ht]
\centering
\includegraphics[scale=0.2]{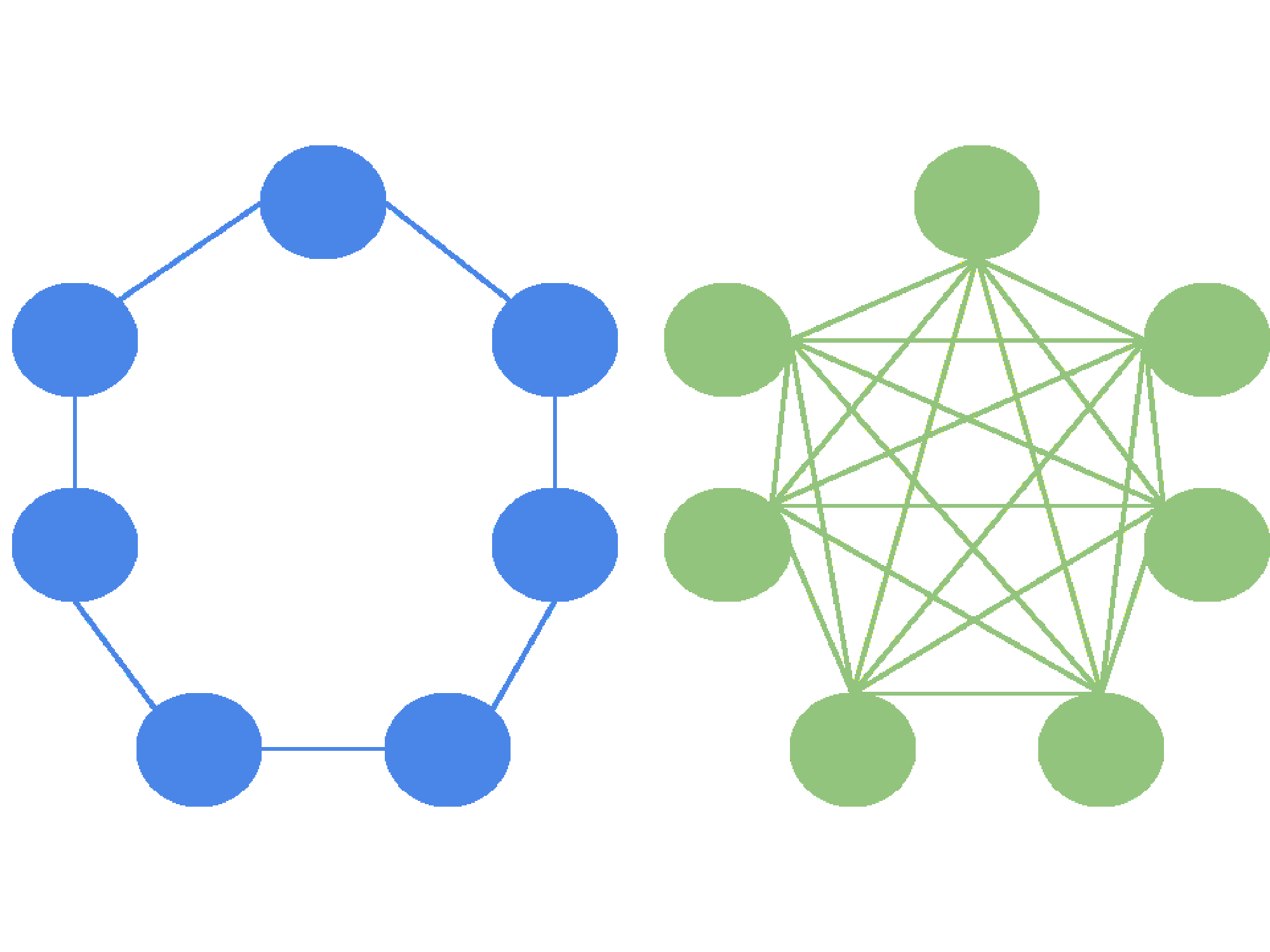}
\caption{Cyclic and complete 7-node Max-Cut instances used as representative test cases.}
\label{fig:cyclic_and_complete_examples}
\end{figure}

Figure \ref{fig:cyclic_and_complete_examples} shows the two 7-node graph topologies used as exemplars. In total, we consider $N \in \{4,7,10\}$ for both cyclic and complete graphs. For each problem/topology we examine three QAOA depths (denoted 1L, 2L, 3L for one, two, and three alternating operator layers, respectively) and both mixer families (RX-only and RX–RY), with and without entanglement inserted in the mixer block.

Our QFI analysis proceeds in three steps:
\begin{enumerate}
    \item \textbf{Random-parameter sampling:} For each setting, 100 random parameter draws in $[0,2\pi)$ are used to probe the nonuniform sensitivity landscape; QFI matrices are averaged to suppress sampling noise.
    \item \textbf{Entanglement-pattern comparison:} We contrast two CNOT-based patterns in the mixer, \emph{cyclic} (nearest-neighbor ring) and \emph{complete} (all-to-all), to isolate the effect of entanglement connectivity on QFI.
    \item \textbf{Entanglement-stage count:} Holding the overall depth fixed (e.g., 3L), we vary the number of entanglement stages (1, 2, or 3) to assess how repeated entangling layers redistribute QFI between diagonal and off-diagonal components. This study is carried out on the 7-node complete graph.
\end{enumerate}

\section{Results and Discussion}

\begin{figure}[!ht] 
\centering 
\includegraphics[scale=0.35]{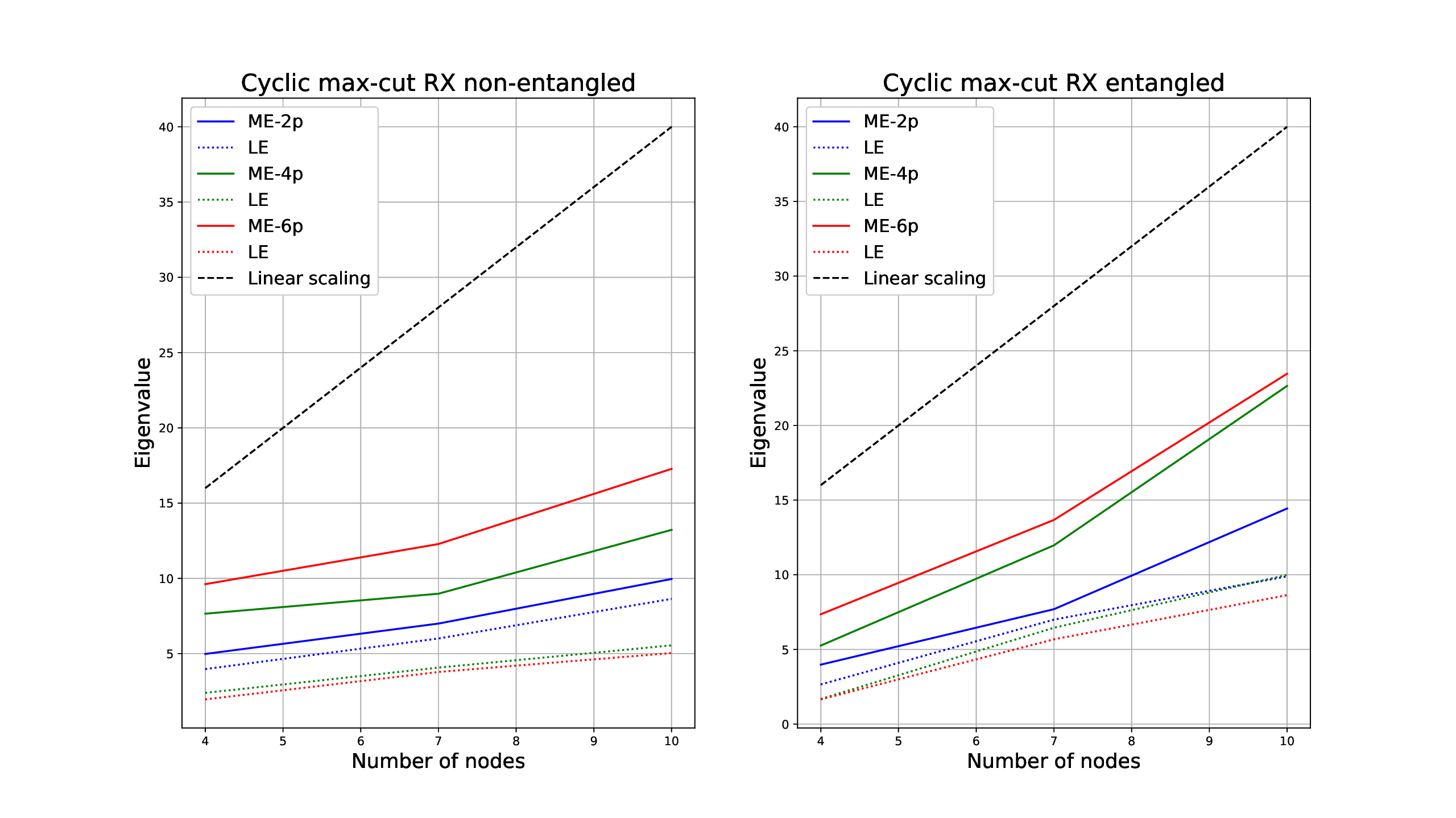} 
\caption{RX-only mixing operators for cyclic max-cut problems with 4, 7, and 10 nodes, using different depths. ME is the maximum eigenvalue of the model and LE is the lowest eigenvalue of the model.} 
\label{fig:rx_cyc_mc_eigenvalues} 
\end{figure}

\begin{figure}[!ht] 
\centering
\includegraphics[scale=0.35]{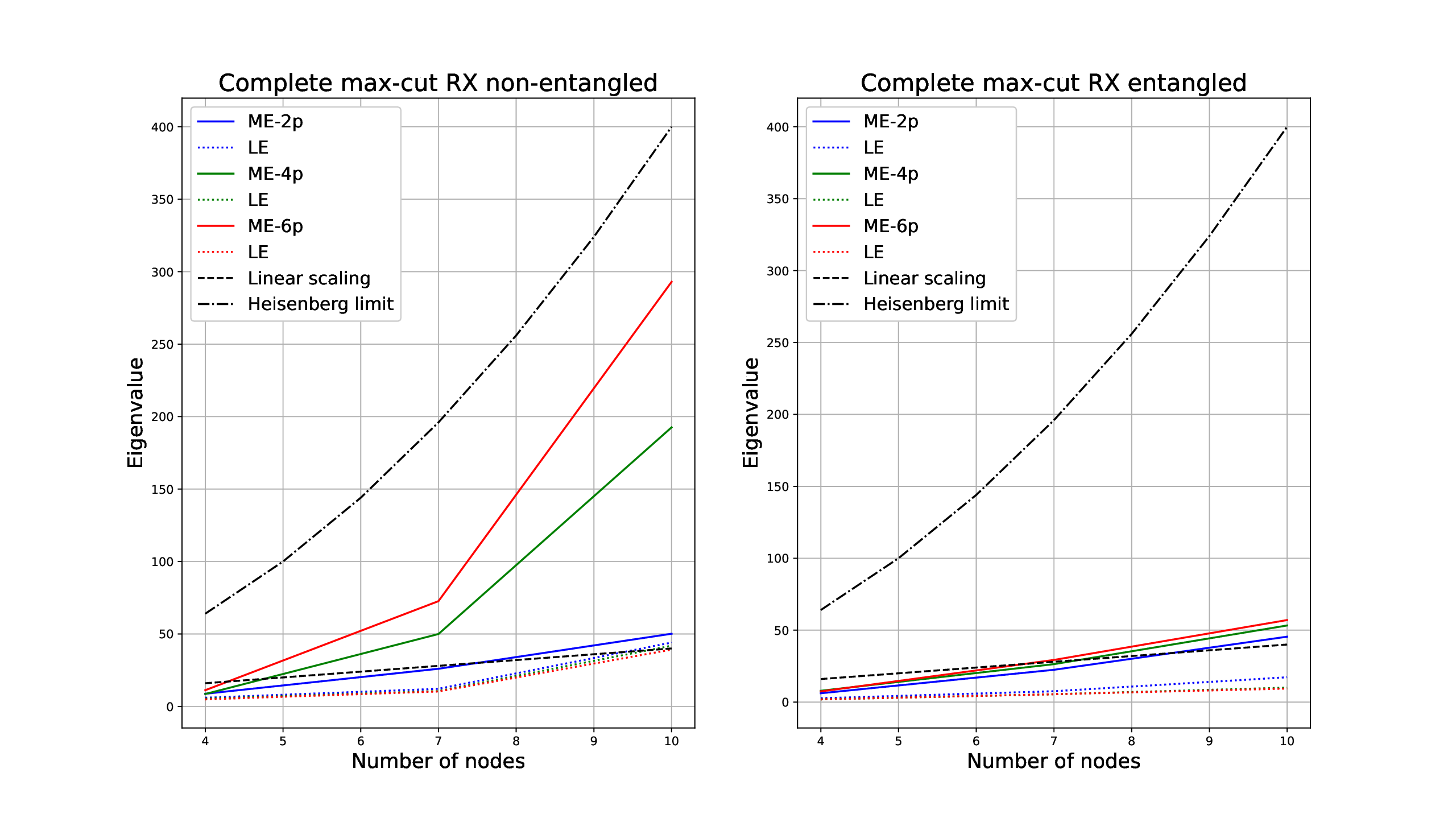} 
\caption{RX-only mixing operators for complete max-cut problems with 4, 7, and 10 nodes, using different depths. ME is the maximum eigenvalue of the model and LE is the lowest eigenvalue of the model.}
\label{fig:rx_com_mc_eigenvalues}
\end{figure}

\begin{figure}[!ht] 
\centering
\includegraphics[scale=0.35]{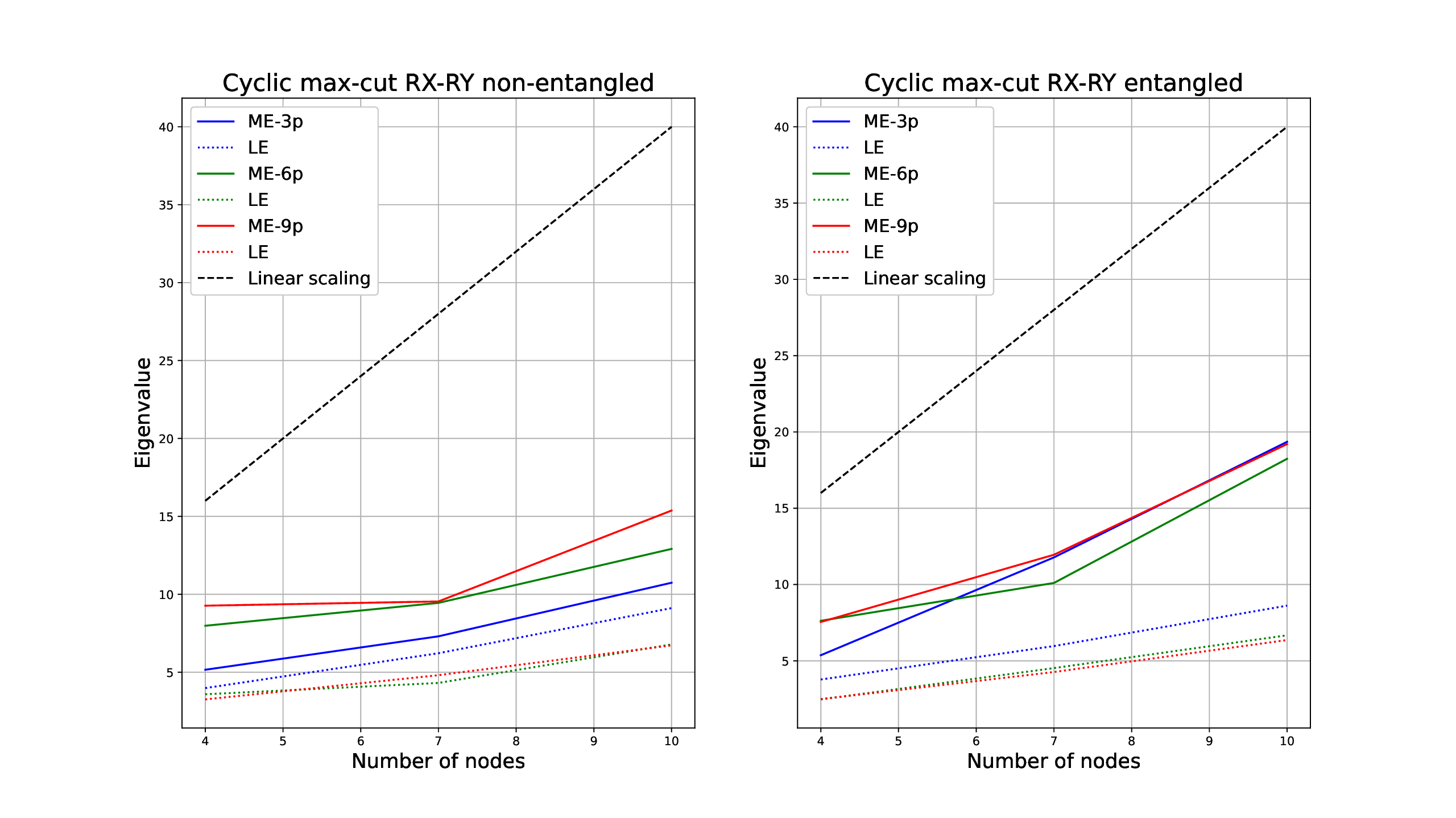}
\caption{RX-RY mixing operators for cyclic max-cut problems with 4, 7, and 10 nodes, using different depths. ME is the maximum eigenvalue of the model and LE is the lowest eigenvalue of the model.}
\label{fig:rxry_cyc_mc_eigenvalues}
\end{figure}

\begin{figure}[!ht] 
\centering
\includegraphics[scale=0.35]{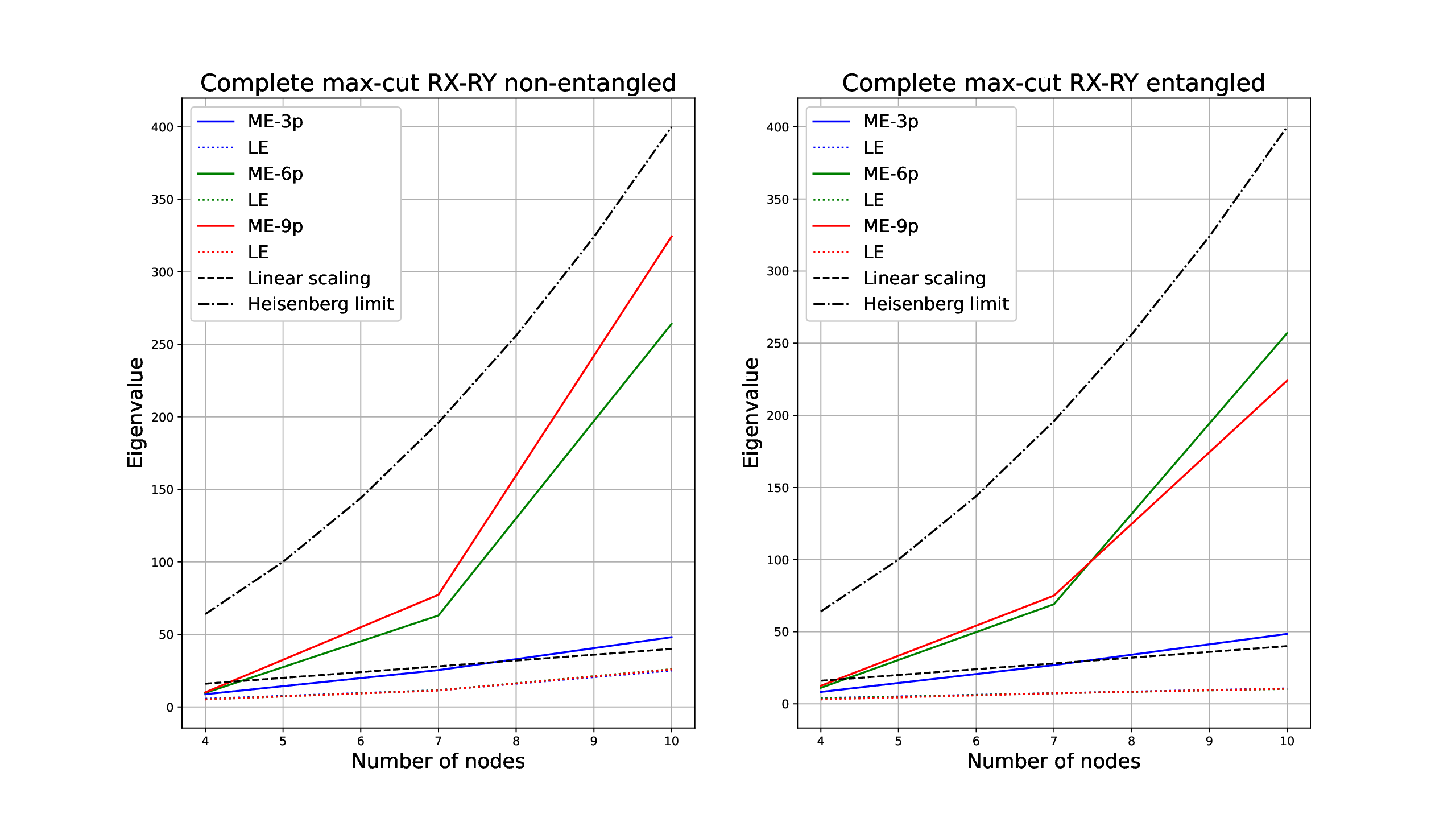}
\caption{RX-RY  mixing operators for complete max-cut problems with 4, 7, and 10 nodes, using different depths. ME is the maximum eigenvalue of the model and LE is the lowest eigenvalue of the model.}
\label{fig:rxry_com_mc_eigenvalues}
\end{figure}

\begin{figure}
     \centering
     \begin{subfigure}[b]{0.49\textwidth}
         \centering
         \includegraphics[width=\textwidth]{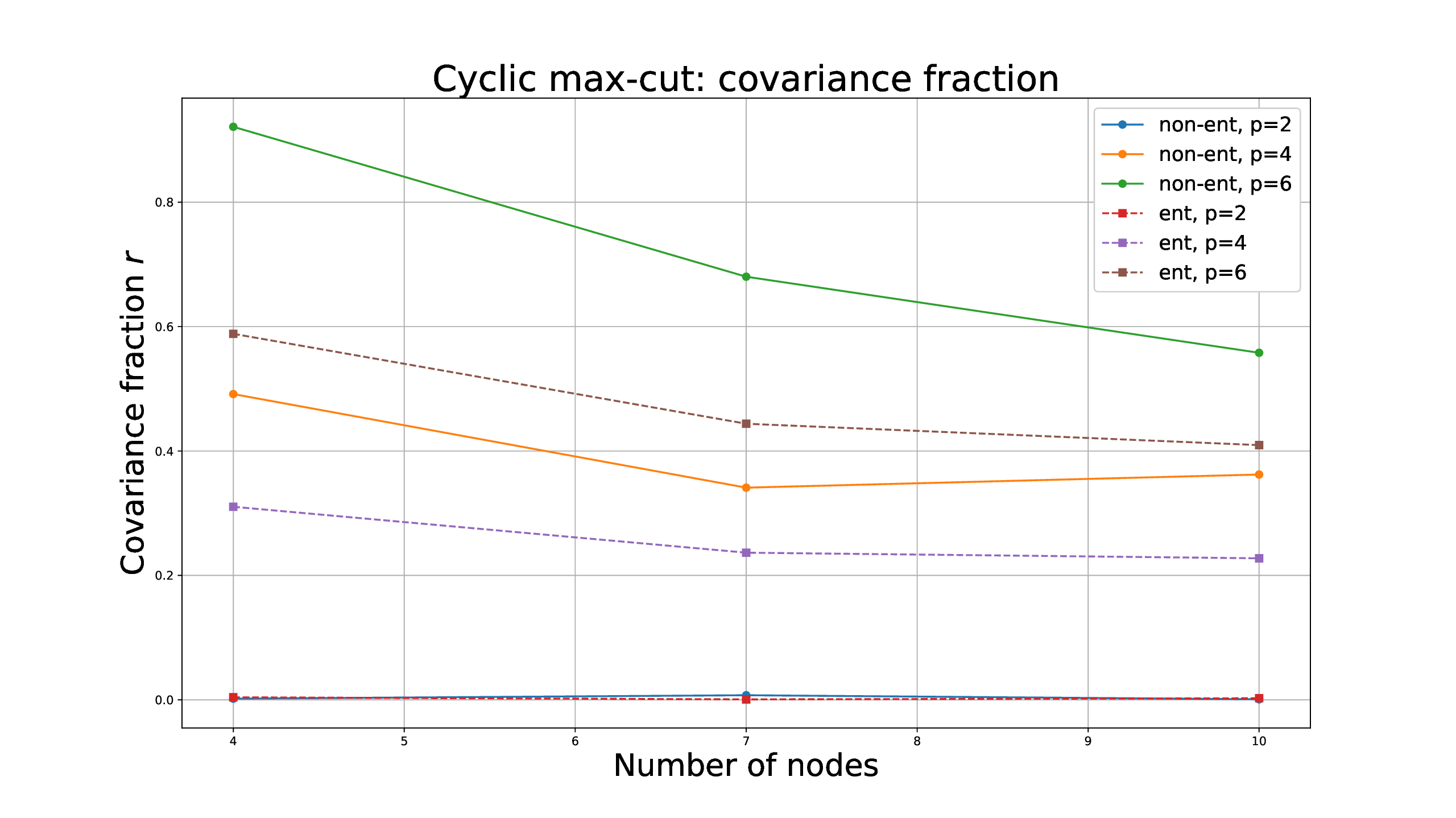}
         \caption{Cyclic configuration.}
         \label{fig:cyc_cov_frac_rx}
     \end{subfigure}
     \hfill
     \begin{subfigure}[b]{0.49\textwidth}
         \centering
         \includegraphics[width=\textwidth]{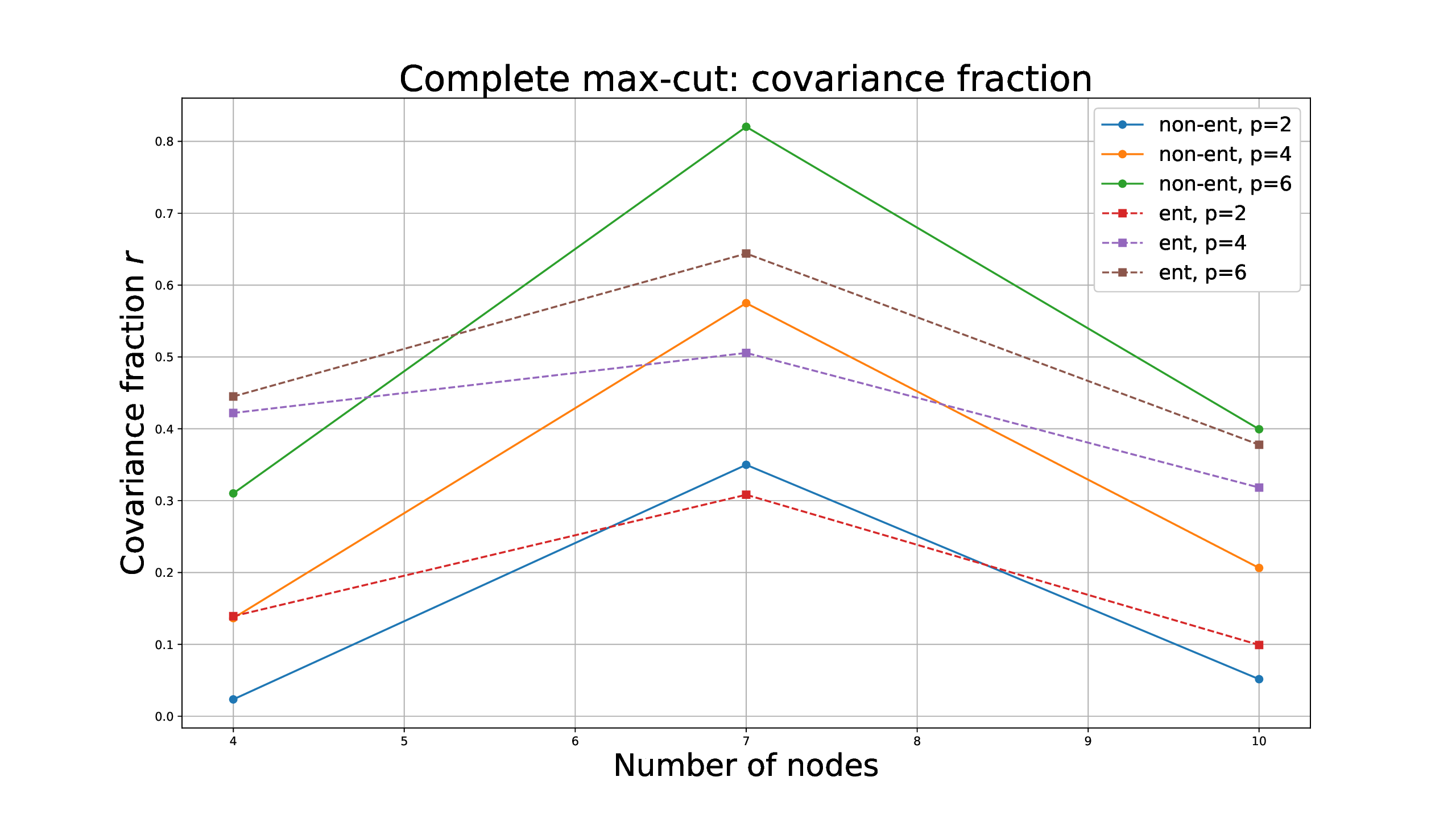}
         \caption{Complete configuration.}
         \label{fig:com_cov_frac_rx}
     \end{subfigure}
    \caption{Covariance fraction for the cyclic and complete max-cut problems with RX-only mixers different number of nodes and different QAOA depths, with and without entanglement stage(s).}
    \label{fig:rx_cov_fractions}
\end{figure}

\begin{figure}
     \centering
     \begin{subfigure}[b]{0.49\textwidth}
         \centering
         \includegraphics[width=\textwidth]{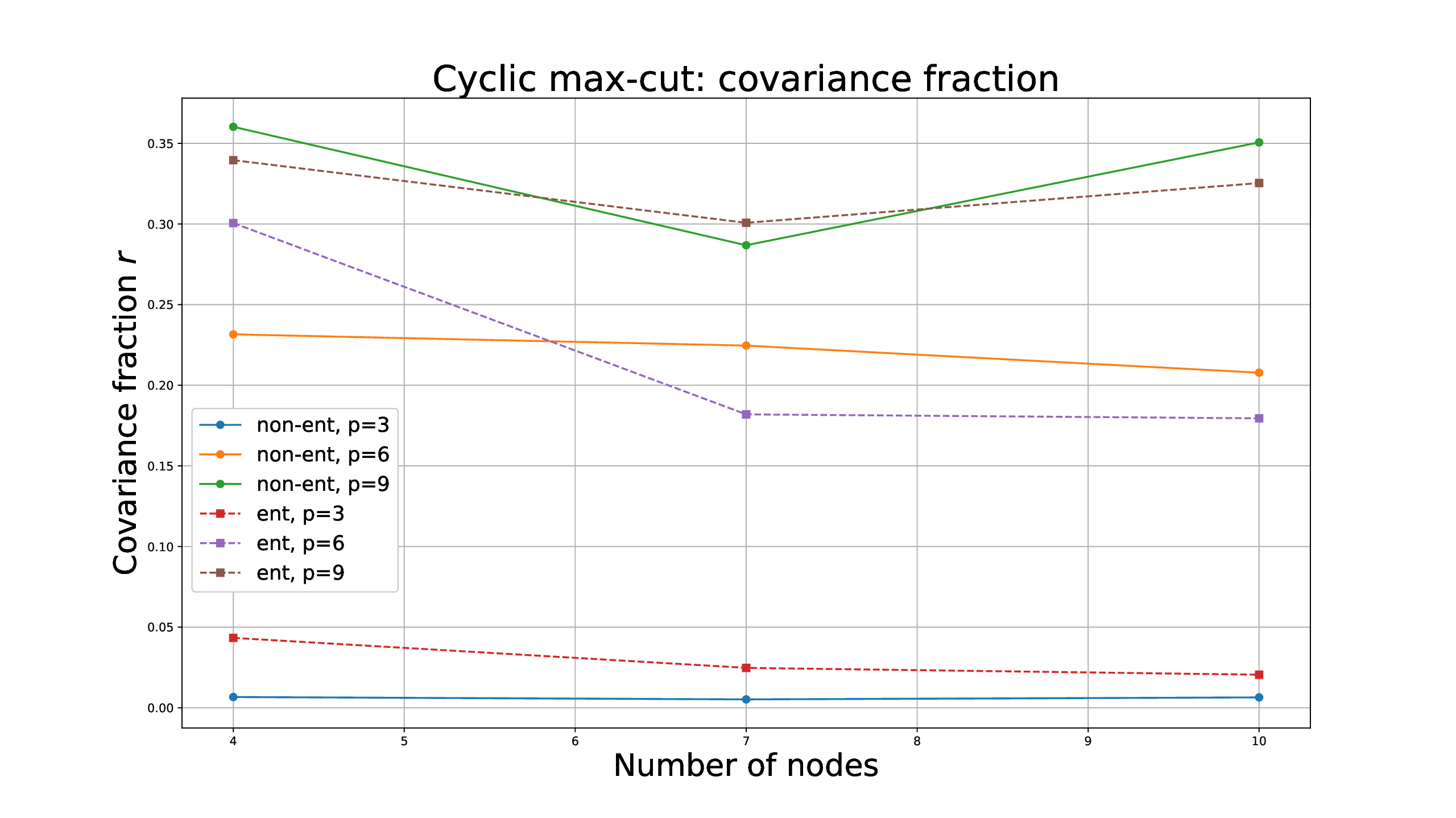}
         \caption{Cyclic configuration.}
         \label{fig:cyc_cov_frac_rx_ry}
     \end{subfigure}
     \hfill
     \begin{subfigure}[b]{0.49\textwidth}
         \centering
         \includegraphics[width=\textwidth]{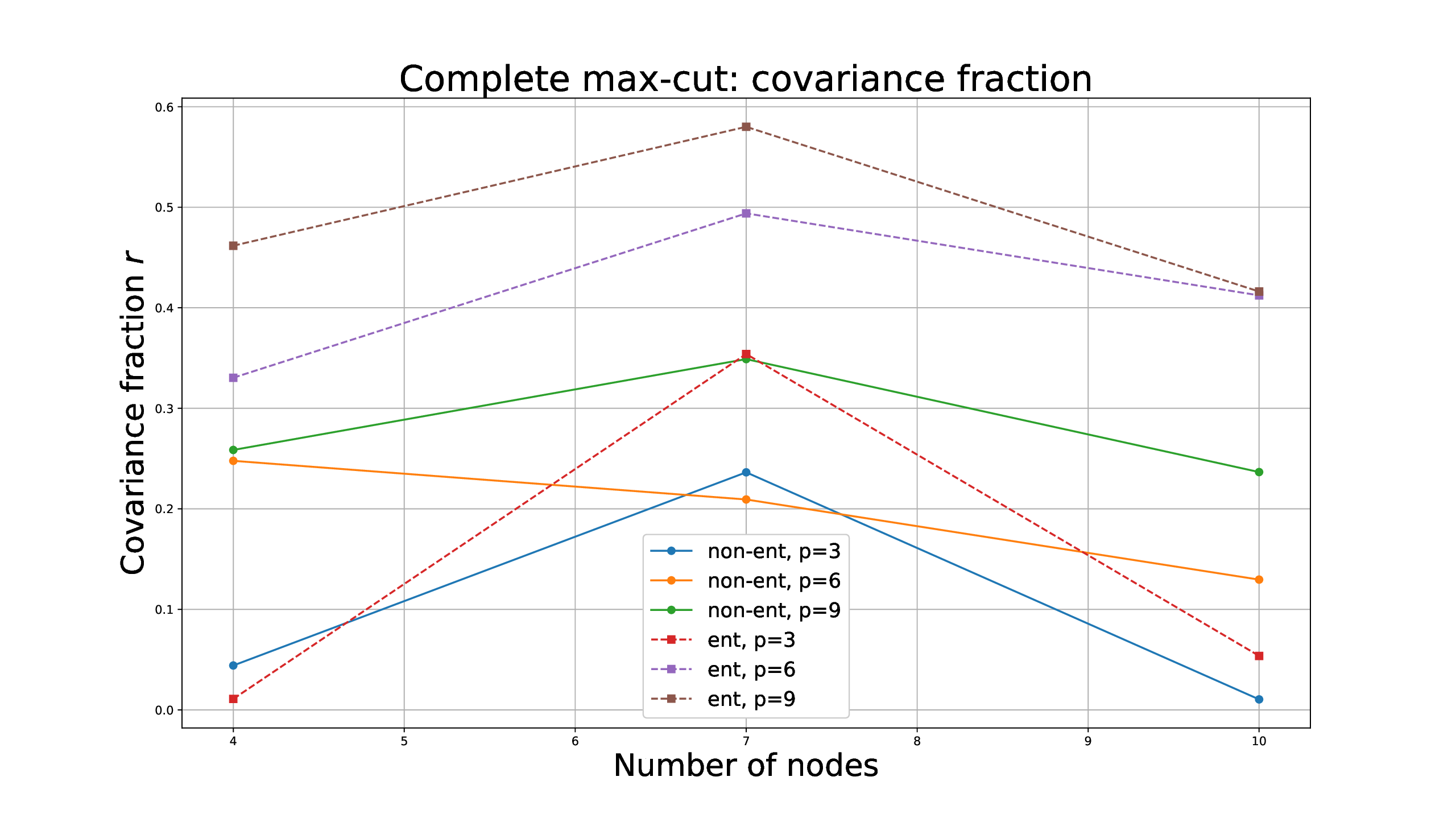}
         \caption{Complete configuration.}
         \label{fig:com_cov_frac_rx_ry}
     \end{subfigure}
    \caption{Covariance fraction for the cyclic and complete max-cut problems with RX-RY mixers different number of nodes and different QAOA depths, with and without entanglement stage(s).}
    \label{fig:rxry_cov_fractions}
\end{figure}

We report Quantum Fisher Information (QFI) for all tested QAOA settings by examining (i) the extremal eigenvalues of the QFI matrix and (ii) the covariance fraction. Figures \ref{fig:rx_cyc_mc_eigenvalues},\ref{fig:rx_com_mc_eigenvalues},\ref{fig:rxry_cyc_mc_eigenvalues}, and \ref{fig:rxry_com_mc_eigenvalues} display the maximum (ME) and lowest (LE) eigenvalues; while Figures \ref{fig:rx_cov_fractions} and \ref{fig:rxry_cov_fractions} show the corresponding covariance fractions.

Across all scenarios, no configuration attains the Heisenberg limit $4N^{2}$; nevertheless, several complete-graph instances exceed the linear bound $4N$. The minimum eigenvalues remain on the order of $1$–$10$, indicating broad spectra. Complete graphs consistently yield larger MEs than cyclic graphs for fixed mixer and depth, reflecting the denser two-body structure of the cost Hamiltonian. RX–RY mixers achieve eigenvalues comparable to, or larger than, RX-only mixers on complete graphs but stay closer to the linear limit on cyclic graphs. In every case the dominant $N^{2}$ term originates from the cost layer; mixer layers primarily redistribute spectral weight rather than set the overall scale. Entanglement stages affect the spectra asymmetrically: for RX-only circuits, adding an entangling layer increases the ME on cyclic graphs but reduces it on complete graphs beyond the first stage, effectively compressing the spectrum. For RX–RY circuits, MEs grow with depth but gains saturate quickly; additional stages produce only marginal or non-monotonic improvements.

The covariance fraction corroborates these trends. For shallow circuits ($p=2$ or $3$ parameters) $r$ is small-$r\lesssim 0.15$ in RX-only and $r\lesssim 0.05$ for RX–RY on cyclic graphs—indicating near-diagonal dominance. Increasing depth drives $r$ upward: in RX-only, non-entangled cyclic instances reach $r\approx 0.9$ at $p=6$ (peaking around $N=7$), whereas entangled variants plateau near $r\approx 0.6$. On complete graphs the peak is slightly lower ($r\approx 0.8$ non-entangled and $\sim 0.64$ entangled) and tapers at $N=10$. RX–RY models display moderate cross-talk: on complete graphs $r$ increases with $p$ and peaks near $N=7$, while cyclic graphs remain uniformly lower ($r\lesssim 0.4$). A non-monotonic pattern appears for complete-entangled RX-RY circuits, where $r$ dips at the second stage and recovers at the third. In all cases, complete-entanglement patterns yield larger $r$ than cyclic-entanglement, and the first entangling stage produces the largest jump; subsequent stages often reduce (RX) or only slightly raise (RX–RY/complete) $r$.

Taken together, these results indicate that circuit depth is the primary driver of both eigenvalue growth and parameter cross-talk, whereas graph density and entanglement range fix their absolute scale. High covariance fractions signal strong parameter coupling and thus motivate natural-gradient or full-metric optimization methods, while low-$r$ regimes remain amenable to diagonal or block-diagonal preconditioners.

\subsection{QFI for 7‑node QAOA with varied entanglement stages}

\begin{figure}
\centering
\includegraphics[scale=0.35]{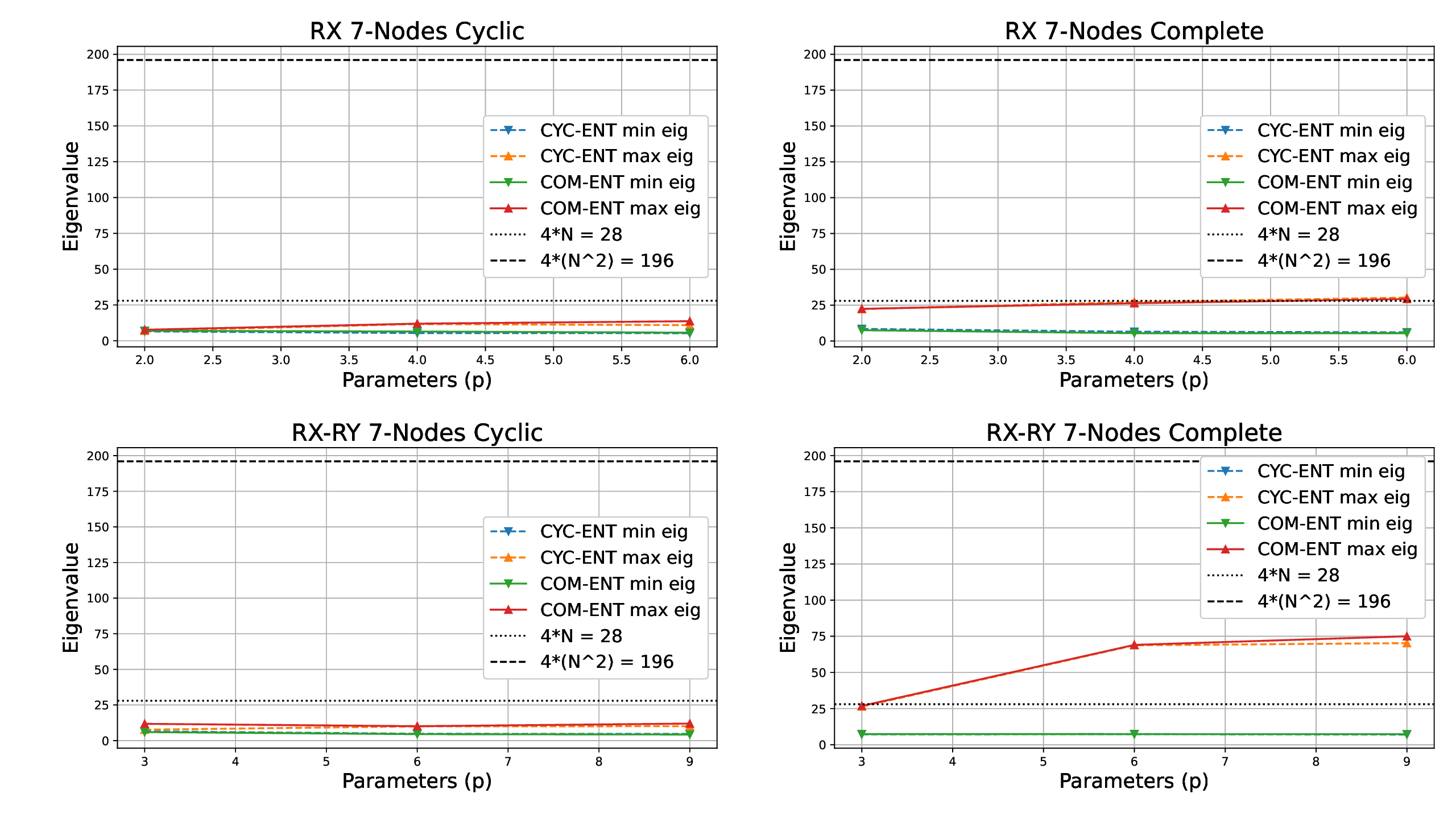}
\caption{Maximum and minimum eigenvalues for cyclic vs. complete entanglement configurations tested for the 7-node problems.}
\label{fig:max_min_eigen_entanglements}
\end{figure}

\begin{figure}
\centering
\includegraphics[scale=0.30]{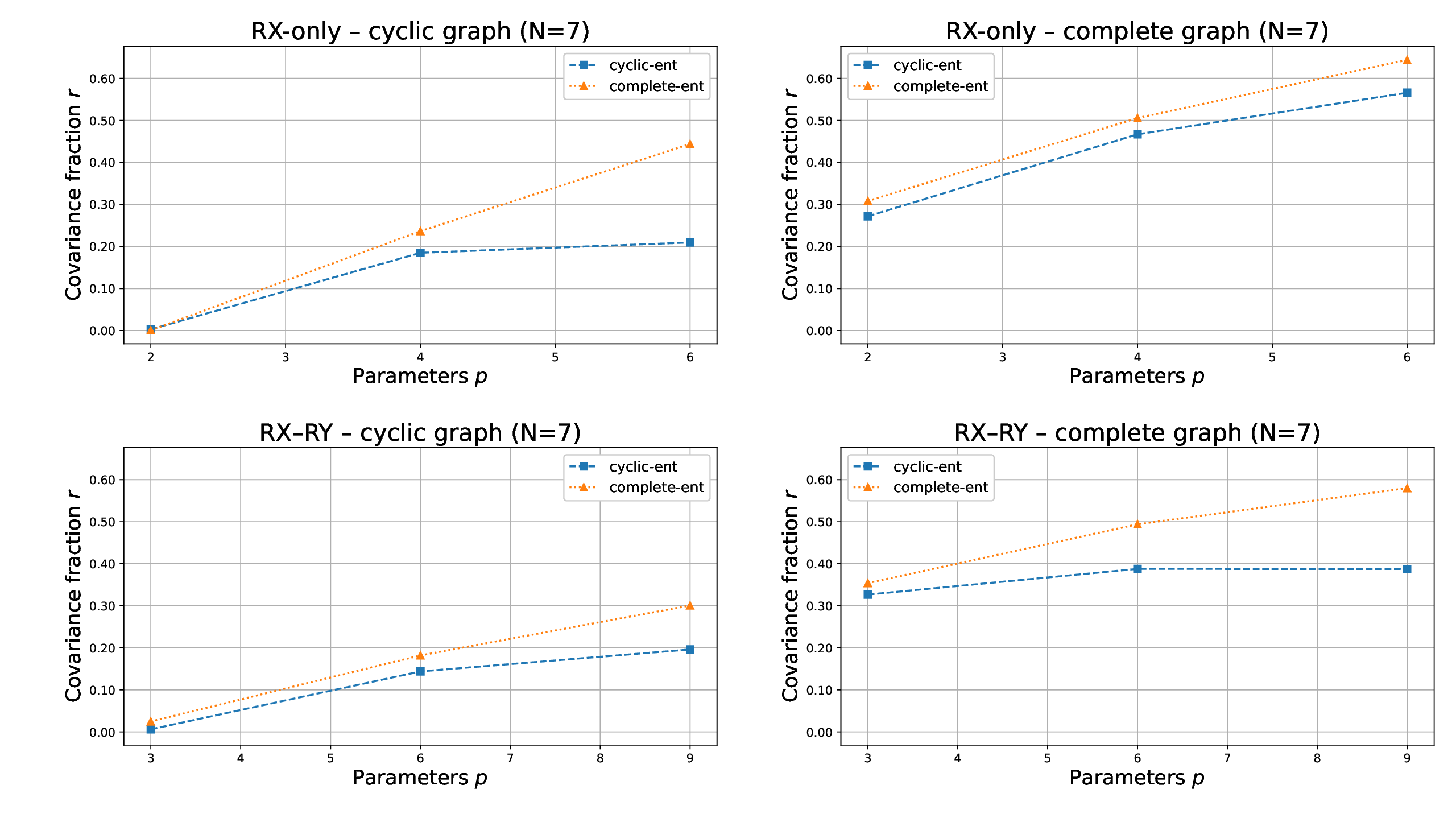}
\caption{Covariance fraction for cyclic vs. complete entanglement configurations tested for the 7-node problems.}
\label{fig:covariance_fraction_cyc_vs_com_ent}
\end{figure}

\begin{figure}
\centering
\includegraphics[scale=0.3]{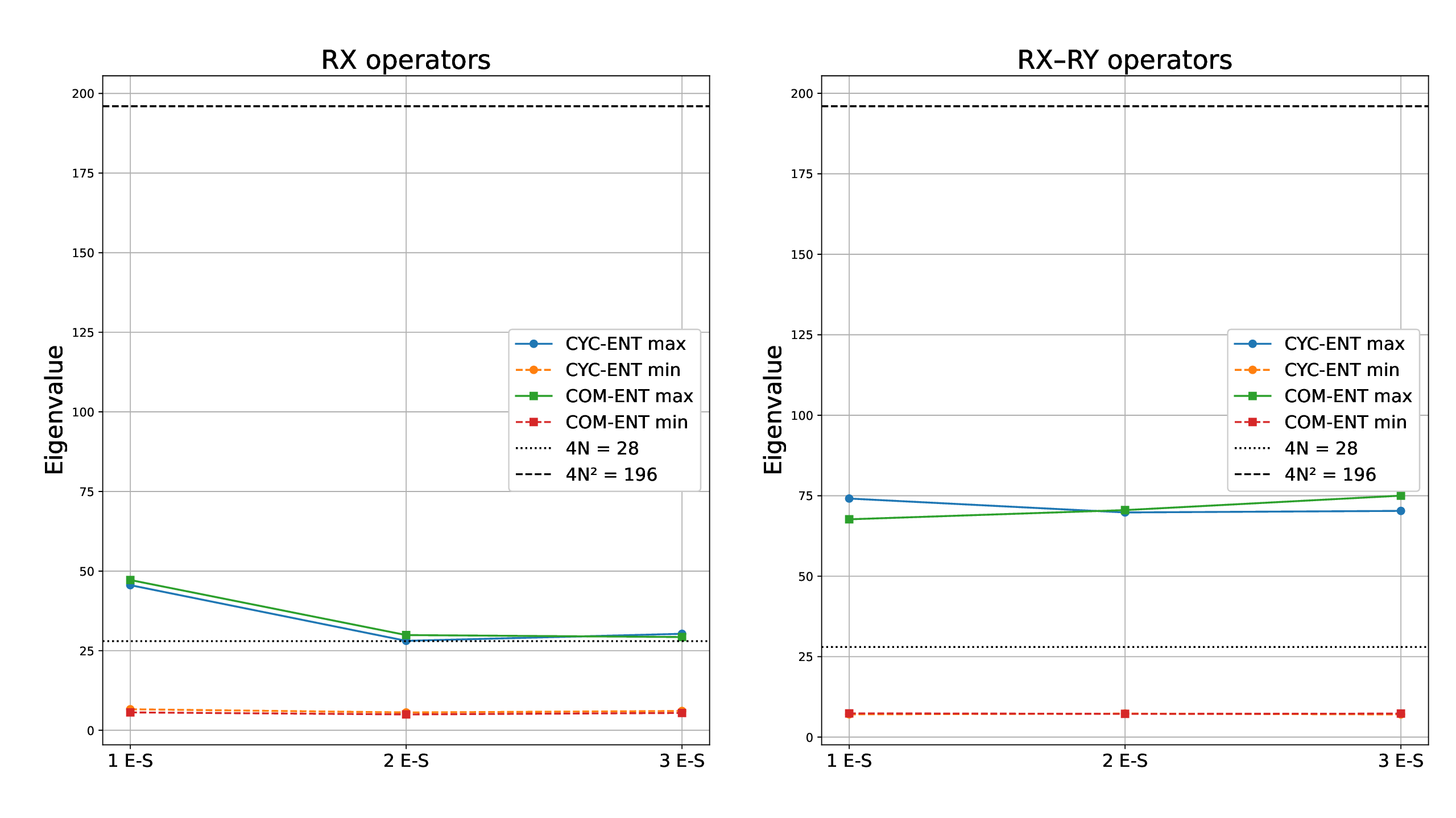}
\caption{Maximum and minimum eigenvalues for the cyclic and complete entanglement configurations at different depths tested for the 7-node complete problem with RX-only and RX-RY mixing operators.}\label{fig:max_min1}
\end{figure}

\begin{figure}
\centering
\includegraphics[scale=0.30]{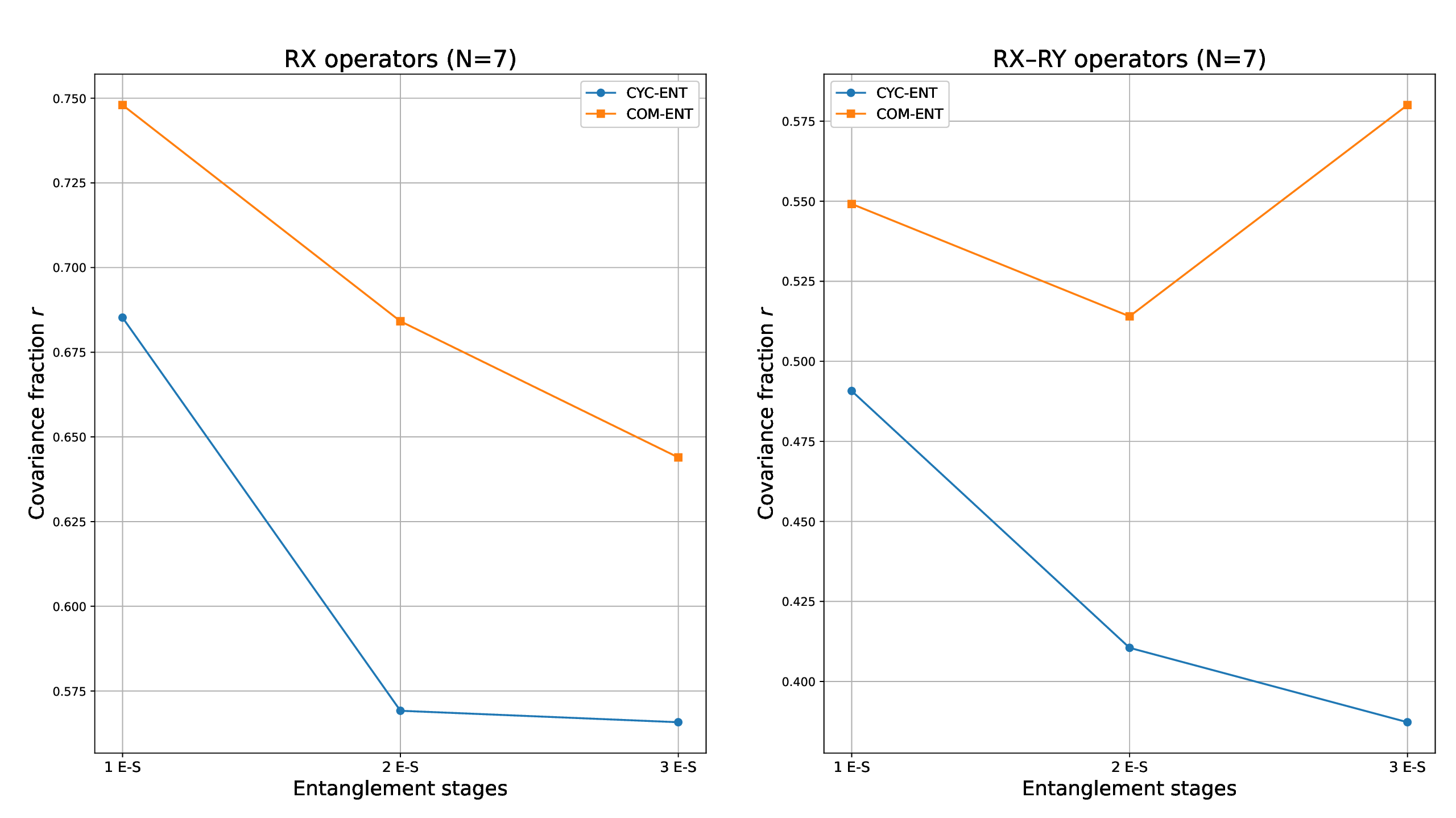}
\caption{Covariance fraction for the cyclic and complete entanglement configurations at different depths tested for the 7-node complete problem with RX-only and RX-RY mixing operators.}\label{fig:cov_frac_max_min1}
\end{figure}

To elucidate the role of entangling layers in shaping QFI, we focus on $N=7$ Max‑Cut instances with both cyclic and complete mixer configurations, and with RX‑only and RX–RY mixers. Unlike our earlier uniform-entanglement setup (where every mixer layer included a complete all‑to‑all CNOT stage), here we also implement a cyclic entanglement pattern—i.e.\ nearest‑neighbour ring CNOTs—and compare it directly against the complete pattern. Figure~\ref{fig:max_min_eigen_entanglements} shows the resulting maximum (ME) and minimum (LE) eigenvalues of the QFI matrix for depths 1L–3L. On cyclic graphs, neither RX nor RX–RY models exceed the linear bound $4N$, and the ME differences between cyclic and complete entanglement are negligible. By contrast, on the complete graph both RX and RX–RY exceed $4N$ for deeper circuits, with RX–RY achieving the largest MEs across all tested configurations—though still well below the Heisenberg limit $4N^2$.
We further quantify cross‑parameter coupling via the covariance fraction $r=\sum_{i\neq j}|F_{ij}|/\sum_i F_{ii}$ plotted in Figure~\ref{fig:covariance_fraction_cyc_vs_com_ent}. In all mixer families, $r$ grows monotonically with depth, reflecting stronger cross‑talk in deeper circuits. Complete-entangled mixers consistently yield larger $r$ than cyclic‑entangled ones, and complete graphs exhibit higher absolute $r$ values. At the highest depths, both RX and RX–RY on complete graphs plateau near $r\approx0.55$–$0.60$, whereas cyclic graphs remain below $r\approx0.45$.
Finally, to isolate the effect of the number of entanglement stages, we fix depth to 3L (nine mixer parameters for RX–RY, six for RX) on the complete 7‑node graph and vary the count of entangling layers from one to three. Figure~\ref{fig:max_min1} reveals that for RX‑only circuits the ME collapses from $\sim45$ at one entangling stage to $\sim30$ at two and three stages (just above $4N=28$), while LE remains $\approx5$. RX–RY circuits sustain higher MEs ($\sim70$–$75$), with a modest monotonic rise for complete entanglement and a shallow dip–recovery for cyclic entanglement. 
The corresponding covariance fractions in Figure~\ref{fig:cov_frac_max_min1} show that RX‑only $r$ falls steadily from $\approx0.75$ (1 stage) to $\approx0.56$ (3 stages), indicating progressive diagonalization, whereas RX–RY exhibits a non‑monotonic pattern (complete: $0.55\to0.51\to0.58$, cyclic: $0.49\to0.42\to0.39$). Thus, while the first entangling layer delivers the largest QFI gain and cross‑talk increase, additional stages offer diminishing and sometimes non‑monotonic returns.
The key findings of the QFI results of the different QAOA and entanglement models are summarized in Table \ref{table1}.
\begin{table}[!ht]
\centering
\caption{Summary of QFI and covariance–fraction findings}\label{table1}
\begin{tabular}{|c|p{10cm}|}
\hline
\textbf{Category} & \textbf{Key insight} \\ \hline
\textbf{Scaling \& bounds} &
No RX or RX–RY configuration attains the Heisenberg limit $4N^2$; the largest eigenvalues remain a few times up from the linear bound $4N$, while the smallest eigenvalues stay $\mathcal{O}(1\text{--}10)$, indicating broad spectral spreads. \\ \hline
\textbf{Graph topology} &
Complete graphs yield systematically larger maximum eigenvalues and higher covariance fractions $r$ than cyclic graphs. For $N=7$, $r_{\rm complete}\approx0.55\text{--}0.60$ versus $r_{\rm cyclic}\approx0.35\text{--}0.45$ (RX–RY) and even lower for RX-only at deep circuits. \\ \hline
\textbf{Mixer family} &
Adding RY rotations raises the maximum eigenvalues modestly but does not substantially alter $r$ compared to the dominant effects of graph topology and entanglement pattern. The cost Hamiltonian remains the principal source of $N^2$ scaling. \\ \hline
\textbf{Covariance fraction} &
Off‑diagonal weight $r$ grows rapidly with depth $p$ from near zero at $p=2$ (e.g.\ $r\lesssim0.15$ for RX, $r\lesssim0.05$ for RX–RY/cyclic) to plateaus at $r\approx0.9$ (RX/cyclic non‑ent), $r\approx0.6$ (RX/cyclic ent), and $r\approx0.8$ (RX/complete non‑ent) or $\sim0.64$ (RX/complete ent). RX–RY/complete shows non‑monotonic behavior ($0.55\to0.51\to0.58$ for $p=3,6,9$). \\ \hline
\textbf{Entanglement depth} &
The first entangling layer delivers the largest increase in both eigenvalues and $r$. Subsequent layers compress RX-only spectra (lower max eigenvalue, smaller $r$) or yield marginal and sometimes non‑monotonic gains in RX–RY. \\ \hline
\textbf{System size $N$} &
Eigenvalues and $r$ rise with $N$ up to $N=7$ then decline slightly at $N=10$, as diagonal contributions begin to dominate in larger systems. \\ \hline
\textbf{Optimization guidance} &
Low $r\lesssim0.2$ indicates near-independent parameters (diagonal preconditioning suffices). High $r\gtrsim0.4$–$0.6$ signals strong cross-talk, recommending natural-gradient or full-metric methods. \\ \hline
\end{tabular}
\end{table}

\subsection{QFI-Driven Parameter Updates in QAOA}

To illustrate how QFI can accelerate QAOA parameter tuning, we present two proof‑of‑concept case studies. First, we analyze the 7‑node complete‑graph Max‑Cut instance at depth 3L with an RX‑only mixer (six parameters; see Fig.~\ref{figrx4}), and second, the 10‑node cyclic‑graph instance at depth 3L with an RX–RY mixer interleaved with entangling layers (nine parameters; see Fig.~\ref{figrxy5}). From each QFI matrix we derive the QFI‑Informed Mutation (QIm) heuristic: at every iteration, parameter $\theta_i$ is chosen for mutation with probability proportional to its normalized diagonal QFI entry $d_i$, and its step size is set proportional to $1-d_i$. Thus, highly sensitive parameters are mutated more often but in smaller increments, while less sensitive parameters undergo larger, less frequent adjustments. This QFI‑driven scheme focuses computational effort on the most informative directions, leading to faster convergence toward high‑quality cuts.  

\begin{itemize}
    \item \textbf{Mutation probability:} Parameters with larger QFI entries—i.e., those to which the circuit is more sensitive—are mutated more frequently:
    \begin{equation}
        P[\text{mutate } i] = d_{i},
    \end{equation}
    where $\vec{d}\in[0,1]^{p}$ with $p = 2\cdot\text{depth}$, and
    \[
        d_{i} = \biggl[\mathrm{diag}\Bigl(\tfrac{\mathcal{F}}{\mathcal{F}_{\max}}\Bigr)\biggr]_{i},
        \quad
        \mathcal{F}_{\max} = \operatorname{Tr}[\mathcal{F}].
    \]
    \item \textbf{Step mutation ($k_{i}$):} Highly sensitive parameters receive smaller adjustments, while less sensitive ones receive larger adjustments:
    \begin{equation}
        \bigl|\Delta \theta_{i}\bigr| = 1 - d_{i}.
    \end{equation}
    \item \textbf{Variance profile:} The per-parameter update variance is
    \begin{equation}
        \mathrm{Var}[\Delta \theta_{i}] = d_{i}\,(1 - d_{i})^{2},
    \end{equation}
    which is maximized when $d_{i} = 0.5$ and vanishes as $d_{i}\to 0$ or $d_{i}\to 1$.
\end{itemize}
For each coordinate $i$, draw two random variables:
\begin{equation}
    m^{(t)}_{i} \sim \mathrm{Bernoulli}(d_{i}), 
    \quad 
    \zeta^{(t)}_{i} \sim \{-1, 1\}\ (\text{Rademacher}).
\end{equation}
Then, independently and in parallel, update each parameter:
\begin{equation}
    \theta^{(t+1)}_{i}
    = \theta^{(t)}_{i}
    + m^{(t)}_{i}\,\zeta^{(t)}_{i}\,(1 - d_{i}),
    \quad
    i = 1, \ldots, p,
\end{equation}
where $p$ is the number of parameters in the QAOA model.
In vector form, using the Hadamard product $\odot$, the update reads:
\begin{equation}
    \vec{\theta}^{(t+1)}
    = \vec{\theta}^{(t)}
    + (1 - \vec{d}) \odot \vec{\zeta}^{(t)} \odot \vec{m}^{(t)}.
\end{equation}

Now, we present statistical results comparing this QIm against two baseline approaches. The first baseline, nonQIm, fixes $d_{i} = 0.5$, so that each parameter is mutated with equal probability during the Bernoulli update, and uses a constant mutation magnitude $s_{m} = 0.01$. The second baseline, the random-restart heuristic (RR), generates fresh random parameters in $[0,\pi)$ for each $\gamma$ and in $[0,2\pi)$ for each $\beta$, also with $s_{m} = 0.01$. Neither baseline exploits any QFI information. We ran each method for 100 independent experiments, each comprising 100 optimization iterations, to average out stochastic effects and evaluate overall performance.
\begin{figure}
     \centering
     \begin{subfigure}[b]{0.49\textwidth}
         \centering
         \includegraphics[width=\textwidth]{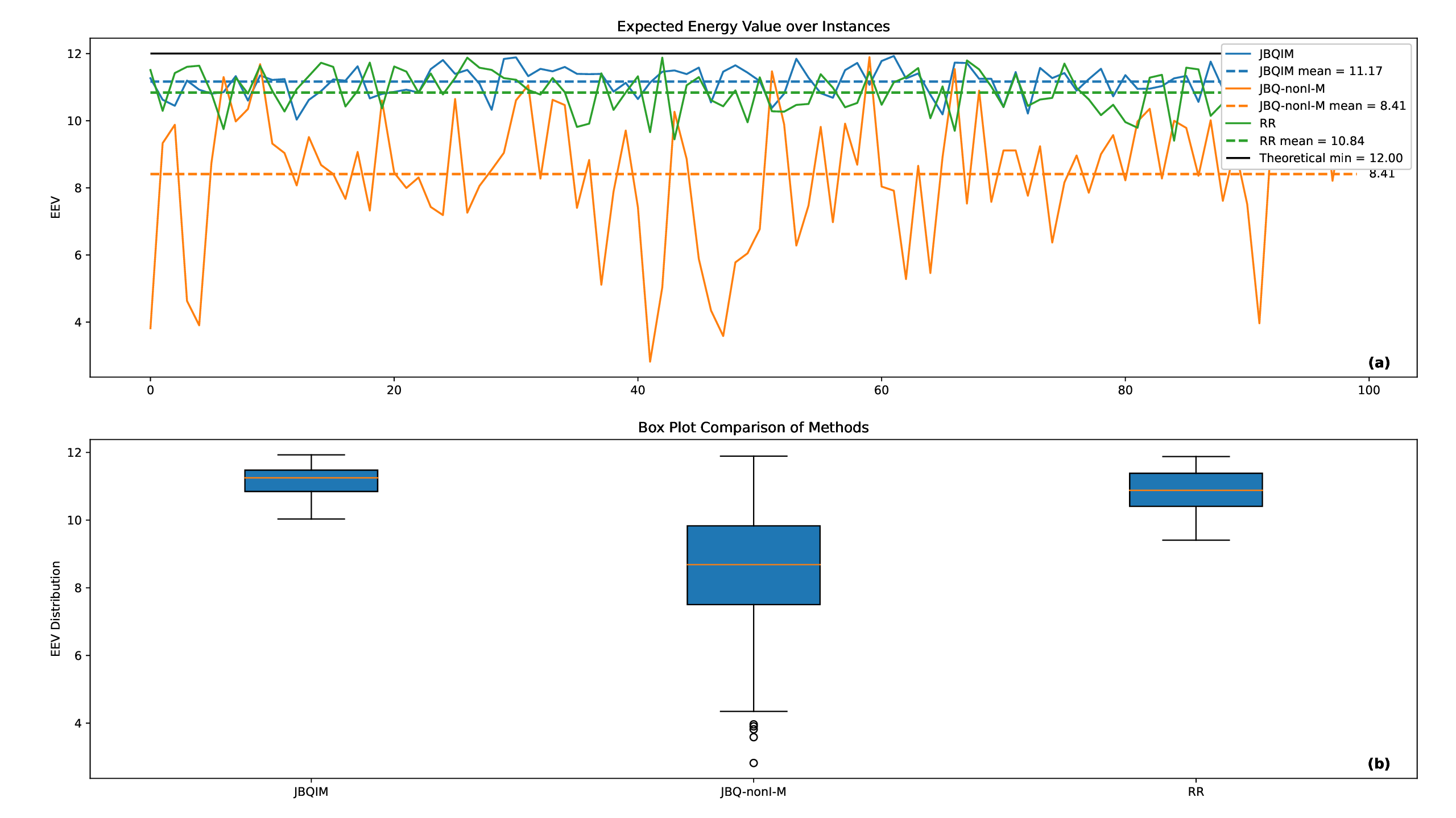}
         \caption{7-node complete‐graph (non‐entangled).}
         \label{fig:7n_com_jbqim}
     \end{subfigure}
     \hfill
     \begin{subfigure}[b]{0.49\textwidth}
         \centering
         \includegraphics[width=\textwidth]{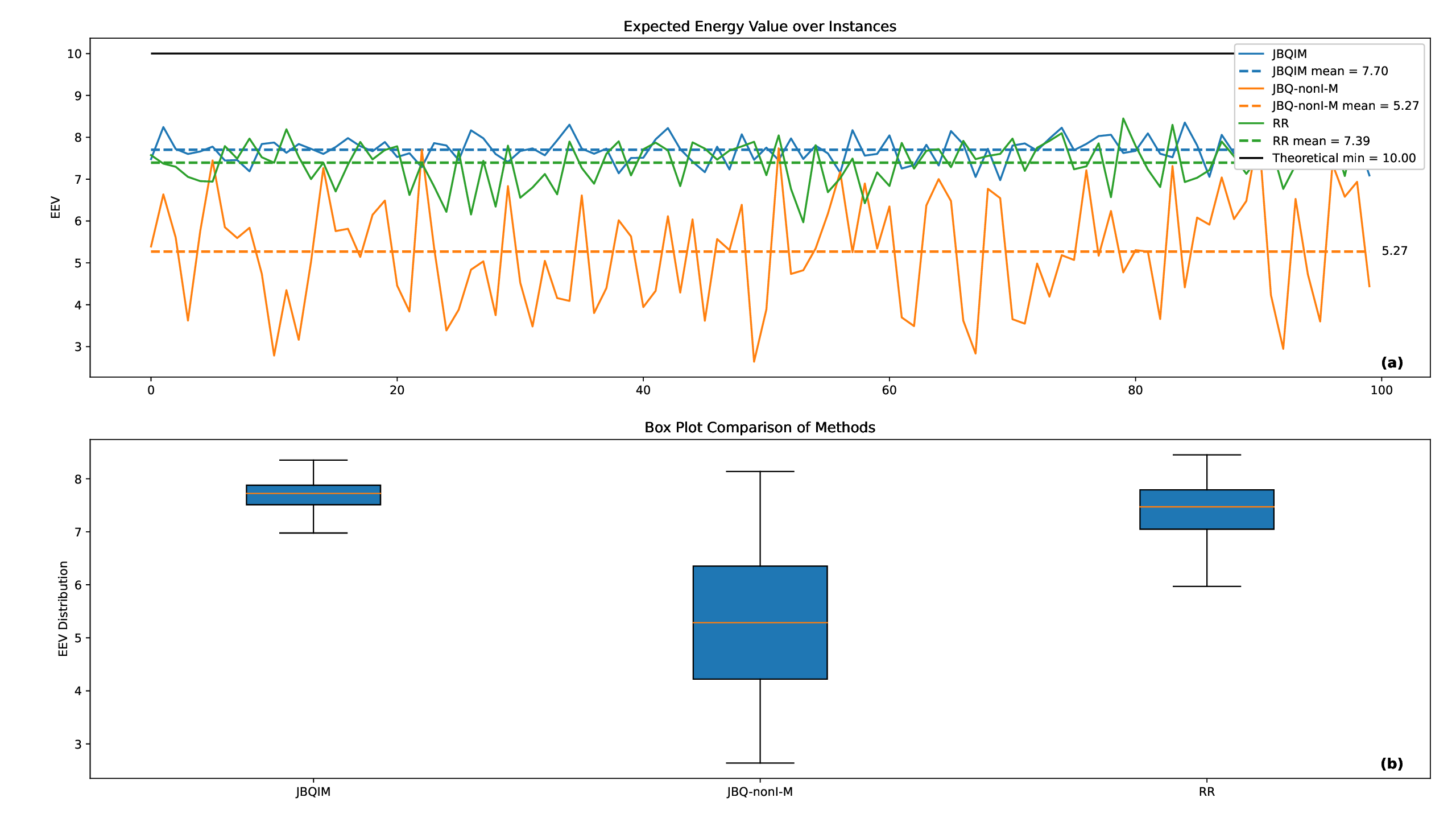}
         \caption{10-node cyclic (entangled).}
         \label{fig:10n_cyc_jbqim}
     \end{subfigure}
    \caption{Results of 100 simulations for QIm, nonQIm, and RR on the (a) 7-node complete‐graph (non‐entangled) and (b) 10‐node cyclic (entangled) Max‐Cut problems. The top panels show the Expected Energy Value (EEV) for each simulation, while the bottom panels present box plots of each model’s mean performance and variance.}
    \label{fig:jbqim_comparison}
\end{figure}
Figure \ref{fig:jbqim_comparison} contrasts our QIm heuristic against two baselines, nonQIm (equal-probability, fixed step size) and RR, over 100 independent runs (each 100 iterations). In both the 7‑node complete‑graph (non‑entangled) case (Fig.\ref{fig:7n_com_jbqim}) and the 10‑node cyclic (entangled) case (Fig.\ref{fig:10n_cyc_jbqim}), QIm consistently achieves higher expected energy values (EEV) and exhibits significantly reduced variance. Median EEV under QIm lies closer to the optimal cut value, while nonQIm and RR show broader spreads and lower central tendencies. These results demonstrate that even an averaged, off‑line QFI matrix can effectively prioritize and scale parameter updates, yielding more reliable convergence than uninformed or random‑restart strategies. Incorporating QFI as a preprocessing guide thus offers a lightweight yet powerful enhancement to classical optimizers in QAOA and other variational algorithms.

\section{Conclusions}

We have shown that the Quantum Fisher Information (QFI) provides a systematic lens into the parameter sensitivity of QAOA applied to Max‐Cut on cyclic and complete graphs. Neither RX‐only nor RX–RY mixers attain Heisenberg‐limit scaling $4N^{2}$, though deeper circuits and complete‐graph topologies can drive some diagonal QFI entries above the linear bound $4N$. 

Entangling layers redistribute Fisher information: unentangled circuits concentrate sensitivity along individual parameters, whereas entangled circuits bolster off‐diagonal (cross‐parameter) correlations, with the first entangling stage delivering the largest effect and additional stages yielding diminishing or non‐monotonic returns. The hybrid RX–RY mixer injects further degrees of freedom that modestly raise maximum QFI values but do not uniformly dominate RX‐only performance. 

Finally, we demonstrated that averaged, precomputed QFI matrices can inform mutation probabilities and step sizes in a simple QFI‐Informed Mutation (QIm) heuristic, substantially improving expected cut performance and reducing variance compared to uninformed or random‐restart baselines. These results highlight QFI both as a diagnostic for circuit design and as a lightweight preconditioning tool for variational parameter optimization, paving the way toward more efficient and noise‐resilient QAOA implementations on NISQ devices.

\noindent
\vskip 5mm
{\large \bf Acknowledgment}: We acknowledge partial support from grants 20240220 and 20240421-SIP-IPN in Mexico. S.H. Dong started this work on the research stay in China.

\bibliographystyle{IEEEtran}
\bibliography{references}

\section{Data availability}

The code and QFI results for the problems studied are publicly available at \url{https://github.com/BrianSarmina/Papers/tree/main/Exploring%20Entanglement%20and%20Parameter%20Sensitivity%20in%20QAOA%20through%20Quantum%20Fisher%20Information}

The QFI matrices used in the JBQIM optimization comparison are provided in Appendix~B.

\appendix
\section{Linear Combination of Unitaries (LCU)} 

Consider a parameterized quantum circuit represented by a unitary operator $U(\vec{\theta})$, where $\vec{\theta}$ is a vector of parameters. This unitary operator acts on an initial state $|\psi_{0}\rangle$, i.e. $|\psi(\vec{\theta})\rangle = U(\vec{\theta})|\psi_{0}\rangle$, its derivative with respect to parameter $\theta_{i}$ is given by
\begin{equation}
    |\partial_{i} \psi \rangle = \frac{\partial }{\partial \theta_{i}} |\psi(\vec{\theta}) \rangle = \left( \frac{\partial U(\vec{\theta}) }{\partial \theta_{i}} \right) |\psi_{0} \rangle.
\end{equation}

Since most parameterized quantum circuits consist of gates that depend on a single parameter (including the gates used in the QAOA models in this paper, such as RZZ, RZ, RX, and RY gates), the derivative of a single-parameter gate $U_{i}(\theta_{i})$ with respect to its parameter $\theta_{i}$ can be expressed as
\begin{equation}
    \frac{\partial U(\vec{\theta}) }{\partial \theta_{i}} = -i G_{i} U_{i}(\theta_{i}) ,
\end{equation}
where $G_{i}$ is the Hermitian generator of the gate $U_{i}(\theta_{i})$. For example, for a rotation gate $R_{\alpha}(\theta_{i}) = e^{-i\theta_{i}\sigma^{\alpha}/2}$ about axis $\alpha = \left\{ x, y, z \right\}$, the generator can be expressed as $G_{i} = \sigma^{\alpha}/2$. This generator $G_{i}$ can be decomposed into its eigenvalues and eigenprojectors. Since $G_{i}$ for Pauli rotations has eigenvalues $\pm \frac{1}{2}$, we can write it as $G_{i} = \sigma^{\alpha}/2 =\left( |+\rangle \langle +| - |-\rangle \langle -| \right)/2$, where $|\pm \rangle$ are the eigenstates of $\sigma_{\alpha}$.

Now, using the decomposition, the derivative becomes
\begin{equation}
    \frac{\partial U_{i}(\theta_{i}) }{\partial \theta_{i}} = -iG_{i}U_{i}(\theta_{i}) = -\frac{i}{2} \left( |+\rangle \langle +| - |-\rangle \langle -| \right) U_{i}(\theta_{i}),
\end{equation}
which can be rewritten as
\begin{equation}
    \frac{\partial U_{i}(\theta_{i}) }{\partial \theta_{i}} = c_{+}U_{+} + c_{-}U_{-} ,
\end{equation}
where $c_{+} = -\frac{1}{2}$, $c_{-} = +\frac{1}{2}$ and $U_{\pm} = |\pm \rangle \langle \pm | U_{i}(\theta_{i})$. This expression shows that the derivative can be represented as a linear combination of the unitaries $U_{\pm}$.

Extending these concepts to a broader perspective, we can define the derivative of a quantum state $|\partial_{i}\psi \rangle = -iU(\vec{\theta})U_{i}^{\dagger}(\theta_{i})G_{i}U_{i}(\theta_{i}) |\psi_{0}\rangle$. If we express $G_{i}$ as a linear combination for generators with eigenvalues $\pm \lambda$, i.e. $G_{i} = \lambda\left( P_{+} - P_{-} \right)$ with $P_{\pm} = |\pm \rangle \langle \pm |$, we can substitute these generators in the derivative getting $|\partial_{i}\psi \rangle = -i\lambda U(\vec{\theta}) U_{i}^{\dagger}(\theta_{i}) \left( P_{+} - P_{-} \right) U_{i}(\theta_{i}) |\psi_{0}\rangle$.

Now, considering the overlaps $\langle \partial_{i} \psi | \partial_{j} \psi \rangle$ terms we have
\begin{equation}
    \langle \partial_{i} \psi | \partial_{j} \psi \rangle = (-i \lambda_{i})(i \lambda_{j}) \langle \psi (\vec{\theta}) | U_{i}^{\dagger}(\theta_{i}) \left( P_{+}^{(i)} - P_{-}^{(i)} \right) U_{i}(\theta_{i}) \cdots
\end{equation}
\begin{equation}
    \cdots U_{j}^{\dagger}(\theta_{j}) \left( P_{+}^{(j)} - P_{-}^{(j)} \right) U_{j}(\theta_{j}) |\psi (\vec{\theta}) \rangle.
\end{equation}

We can simplify the equation from above as
$\langle \partial_{i} \psi | \partial_{j} \psi \rangle =  \lambda_{i} \lambda_{j} \langle \psi (\vec{\theta}) | O_{ij} |\psi (\vec{\theta}) \rangle$, where $O_{ij}$ is an observable constructed from the projectors and unitaries.

For the phase fix term, the partial derivative can be expressed as
\begin{equation}
    \langle \partial_{i} \psi | \psi \rangle = -i \lambda_{i} \langle \psi (\vec{\theta}) | U_{i}^{\dagger}(\theta_{i}) \left( P_{+}^{(i)} - P_{-}^{(i)} \right) U_{i}(\theta_{i})|\psi (\vec{\theta}) \rangle.
\end{equation}

Then, combining the development above, the QFI matrix elements are computed as
\begin{equation}
    F_{ij} = 4 Re\left[  \lambda_{i} \lambda_{j} \langle \psi (\vec{\theta}) | O_{ij} |\psi (\vec{\theta}) \rangle -  \lambda_{i} \lambda_{j} \langle \psi (\vec{\theta}) | O_{i} |\psi (\vec{\theta}) \rangle \langle \psi (\vec{\theta}) | O_{j} |\psi (\vec{\theta}) \rangle \right] 
\end{equation}
being $O_{ij}$, $O_{i}$ and $O_{j}$ observables derived from the LCU expansion, and $\lambda_{i}, \lambda_{j} = \frac{1}{2}$ for Pauli rotation gates.

\section{QFI matrices for JBQIM}

In this appendix section, we present the QFI matrices used for the JBQIM, JBQ-nonI-M and RR comparison in the Results section.

\begin{figure}[]
\centering
\includegraphics[scale=0.40]{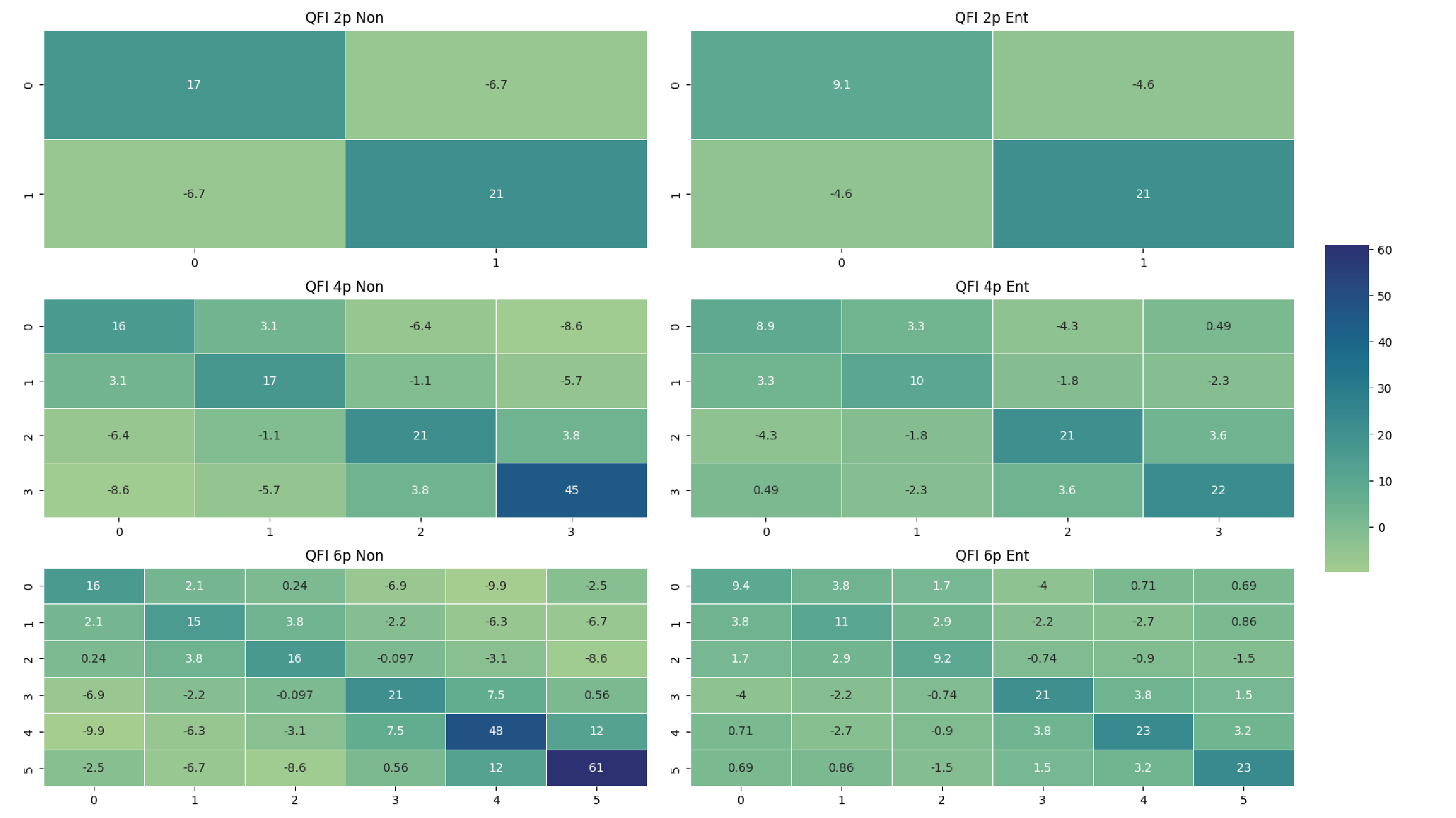}
\caption{QFI matrices for the 7-node max-cut complete configuration problem with RX mixing operators.}
\label{figrx4}
\end{figure}

\begin{figure}[]
\centering
\includegraphics[scale=0.40]{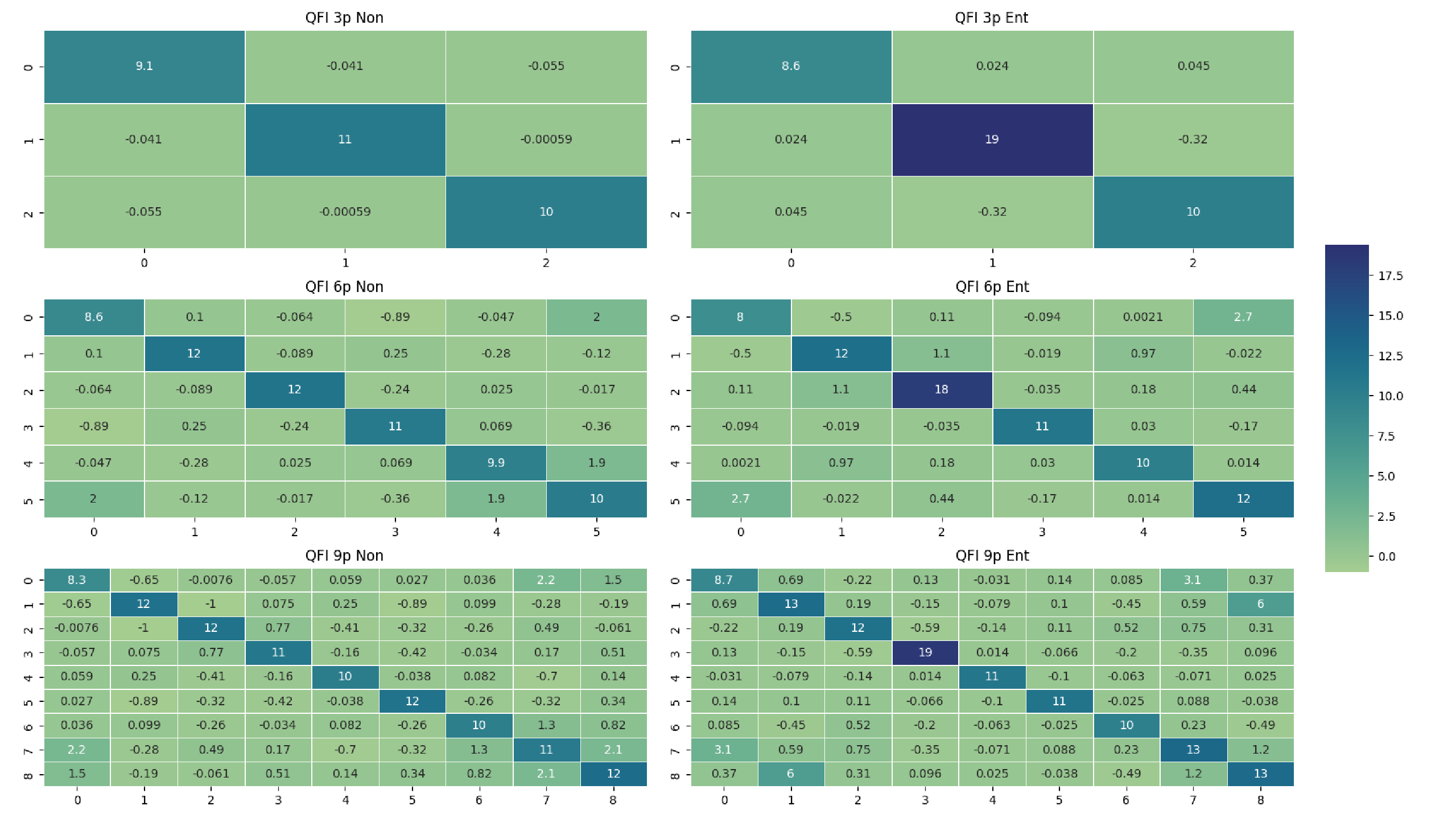}
\caption{QFI matrix for the 10-node max-cut cyclic configuration problem with  RX-RY  mixing operators.}
\label{figrxy5}
\end{figure}

\end{document}